\documentclass[12pt]{article}

\usepackage{comment,rotating}
\usepackage[hang,bf]{caption}
\usepackage{amsmath,amssymb}
\usepackage{color,xcolor,graphicx,graphics,booktabs}
\usepackage{array} 
\usepackage{paralist} 
\usepackage{verbatim} 
\usepackage{subfig} 
\usepackage{multirow}
\usepackage{soul,hyperref}
\usepackage{natbib}
\usepackage{tikz}
\usetikzlibrary{shapes}

\usepackage{titlesec}

\titlespacing\section{0pt}{12pt plus 4pt minus 2pt}{3pt plus 2pt minus 2pt}
\titlespacing\subsection{0pt}{12pt plus 4pt minus 2pt}{3pt plus 2pt minus 2pt}
\titlespacing\subsubsection{0pt}{12pt plus 4pt minus 2pt}{0pt plus 2pt minus 2pt}

\usepackage[belowskip=-5pt,aboveskip=0pt]{caption}

\newcounter{alphnum}

\newcounter{appendix}

\newcounter{Romnum}

\definecolor{lightblue}{RGB}{173,216,230}

\definecolor{rose}{RGB}{255,228,225}

\definecolor{lavender}{RGB}{230,230,250}

\definecolor{lightcyan}{RGB}{224,255,255}
 
\newcommand{\circlerose}{\raisebox{0pt}{\tikz{\filldraw[rose,very thin,line width = 2.0pt](2.mm,0) circle (0.1cm)}}}   

\newcommand{\circlecyan}{\raisebox{0pt}{\tikz{\filldraw[lightcyan,very thin,line width = 2.0pt](2.mm,0) circle (0.1cm)}}}   

\newcommand{\circleblue}{\raisebox{0pt}{\tikz{\filldraw[lightblue,solid,line width = 2.0pt](2.mm,0) circle (0.1cm)}}}   

\newcommand{\circlelav}{\raisebox{0pt}{\tikz{\filldraw[lavender,solid,line width = 2.0pt](2.mm,0) circle (0.1cm)}}}

\newcommand{\nc}{\newcommand}
\nc{\ntitle}[1]{
 \begin{center}
   \fbox{\textbf{\Large #1}}
  \end{center}         }

\nc{\tred}[1]{\textcolor[rgb]{1.00,0.00,0.00}{#1}}
\nc{\tb}[1]{\textcolor[rgb]{0.00,0.00,0.00}{#1}}

\nc{\slideline}{\smallskip \hrule\hrule \smallskip}
\nc{\stitle}[1]{
\textbf{\large #1}
\slideline
\slideline
          }

\nc{\nn}{\nonumber}
\nc{\fns}{\footnotesize}

\nc{\revisionline}{\vspace{.1in} \today \vspace{.1in} \hrule\hrule\hrule\vspace{.1in}}
\nc{\newpp}{\vspace{.1in} \noindent}

\nc{\wh}{\widehat}

\nc{\Ef}{ {\rm E}_{\infty} }
\nc{\Ex}{ {\rm E} }
\nc{\Ec}{ {\rm E}_1 }
\nc{\Pf}{ {\rm P}_{\infty} }
\nc{\Pc}{ {\rm P}_{1} }
\nc{\Prb}{ {\rm P} }
\nc{\sd}{\pm \hat{\sigma} }

\nc{\indep}{{\, \perp \! \! \! \perp  \,} }
\nc{\tsps}{^{ {\rm T} } }

\nc{\pu}{\pi_{\rm U}}
\nc{\pbi}{\pi_{\rm B}}
\nc{\pnb}{\pi_{\rm NB}}
\nc{\prp}{\propto}
\nc{\pr}{ {\rm pr} }

\nc{\al}{\alpha}
\nc{\dl}{\delta}
\nc{\la}{\lambda}
\nc{\om}{\omega}
\nc{\vep}{\varepsilon}

\nc{\snf}{\sum_{n=1}^{\infty}}
\nc{\skf}{\sum_{k=1}^{\infty}}
\nc{\sner}{\sum_{n=1}^{86}}
\nc{\sjn}{\sum_{j=1}^{n}}
\nc{\skn}{\sum_{k=1}^{n}}
\nc{\sumim}{\sum_{i=1}^m}
\nc{\sumjn}{\sum_{j=1}^n}
\nc{\sumlL}{\sum_{l=1}^{L}}
\nc{\sumL}{\sum_{l=1}^{L}}
\nc{\sumkK}{\sum_{k=1}^{K_i}}
\nc{\sumrR}{\sum_{r=1}^R}

\nc{\hivp}{\sum_{ {\rm HIV}^+ } }
\nc{\sumiN}{ \sum_{i=1}^N }
\nc{\summM}{ \sum_{m=1}^M }
\nc{\sumjM}{ \sum_{j=1}^M }

\nc{\lsq}{\left[}
\nc{\rsq}{\right]}
\nc{\lbc}{\left \{ }
\nc{\rbc}{\right \} }
\nc{\lp}{\left(}
\nc{\rp}{\right)}

\nc{\imp}{\Rightarrow}
\nc{\lbf}{\lim_{b \rightarrow \infty}}
\nc{\limNinf}{\lim_{N \rightarrow \infty}}
\nc{\limminf}{\lim_{m \rightarrow \infty}}
\nc{\limninf}{\lim_{n \rightarrow \infty}}
\nc{\convd}{\stackrel{D}{\longrightarrow}}
\nc{\convp}{\stackrel{P}{\longrightarrow}}
\nc{\eqd}{\stackrel{{\EuScript D}}{=}}

\nc{\trans}{^{\text T}}
\nc{\ol}{\overline}
\nc{\logit}{\text{logit}\,}

\nc{\rl}{ {\rm {\bf R} } }
\nc{\zah}{ {\rm {\bf Z} } }

\nc{\lkn}{\Lambda^n_k}
\nc{\stp}{ {\cal C}_b }
\nc{\istp}{ {\cal I}_A }
\nc{\snb}{S_{N_b}}
\nc{\stb}{S_{T_b}}
\nc{\ixlog}{I_{ \{ 0 \leq x \leq \log \al \} } }
\nc{\iulog}{I_{ \{ 0 \leq u  \leq \log \al \} } }
\nc{\rgn}{ \Upsilon_n }
\nc{\var}{{\rm var}}
\nc{\cov}{{\rm cov}}
\nc{\corr}{{\rm corr}}
\nc{\dpl}{\partial}
\nc{\half}{ {\textstyle \frac{1}{2}} }
\nc{\tr}{{\rm trace}}
\nc{\real}{\mathbb{R}}
\nc{\bbC}{\mathbb{C}}

\def\boxit#1{\vbox{\hrule\hbox{\vrule\kern6pt\vbox{\kern6pt#1\kern6pt}\kern6pt\vrule}\hrule}}

\nc{\calb}{ {\cal B} }
\nc{\calc}{ {\cal C} }
\nc{\bcalc}{ \mbox{\boldmath{${\cal C}$}}}
\nc{\cald}{ {\cal D} }
\nc{\cale}{ {\cal E} }
\nc{\cali}{ {\cal I} }
\nc{\call}{ {\cal L} }
\nc{\calm}{ {\cal M} }
\nc{\caln}{ {\cal N} }
\nc{\cals}{ {\cal S} }
\nc{\calo}{ {\cal O} }
\nc{\bcalo}{ \mbox{\boldmath{${\cal O}$}}}
\nc{\calt}{ {\cal T} }
\nc{\calv}{ {\cal V} }
\nc{\bcalu}{ \mbox{\boldmath{${\cal U}$}}}
\nc{\calu}{ {\cal U} }
\nc{\calw}{ {\cal W} }
\nc{\calx}{ {\cal X} }

\nc{\sca}{ {\EuScript A} }
\nc{\scb}{ {\EuScript B} }
\nc{\scc}{ {\EuScript C} }
\nc{\scd}{ {\EuScript D} }
\nc{\sce}{ {\EuScript E} }
\nc{\scf}{ {\EuScript F} }
\nc{\scF}{ {\EuScript f} }
\nc{\scg}{ {\EuScript G} }
\nc{\sch}{ {\EuScript H} }
\nc{\sci}{ {\EuScript I} }
\nc{\scj}{ {\EuScript J} }
\nc{\sck}{ {\EuScript K} }
\nc{\scl}{ {\EuScript L} }
\nc{\sclic}{ \scl_i^{\rm c} }
\nc{\scm}{ {\EuScript M} }
\nc{\scn}{ {\EuScript N} }
\nc{\sco}{ {\EuScript O} }
\nc{\scp}{ {\EuScript P} }
\nc{\scq}{ {\EuScript Q} }
\nc{\scr}{ {\EuScript R} }
\nc{\scs}{ {\EuScript S} }
\nc{\sct}{ {\EuScript T} }
\nc{\scu}{ {\EuScript U} }
\nc{\scv}{ {\EuScript V} }
\nc{\scw}{ {\EuScript W} }
\nc{\scx}{ {\EuScript X} }
\nc{\scy}{ {\EuScript Y} }
\nc{\scz}{ {\EuScript Z} }
\nc{\scxo}{ {\EuScript X}_{\rm obs} }
\nc{\Xobs}{ \pmb{\scx}_{\rm obs} }
\nc{\Xcom}{ \pmb{\scx} }
\nc{\Xmis}{ \pmb{\scx}_{\rm mis} }

\nc{\bsci}{ \mbox{\boldmath{$\sci$}}}
\nc{\bscj}{ \mbox{\boldmath{$\scj$}}}

\nc{\sumlic}{\sum_{l \in sclic}}

\nc{\scyo}{ {\EuScript Y}_{\rm obs} }

\nc{\bga}{\begin{array}{c}}
\nc{\ena}{\end{array}}

\nc{\mhat}{ {\hat{p}}_M }
\nc{\fhat}{ {\hat{p}}_F }
\nc{\ph} { \hat{p} }

\nc{\ta}{ {\tilde{a}} }
\nc{\tc}{ {\tilde{c}} }

\nc{\bal}{\mbox{\boldmath{$\alpha$}}}
\nc{\balpha}{\mbox{\boldmath{$\alpha$}}}
\nc{\bone}{\mbox{\boldmath{$1$}}}
\nc{\bbet}{\mbox{\boldmath{$\beta$}}}
\nc{\bbeta}{\mbox{\boldmath{$\beta$}}}
\nc{\bDel}{\mbox{\boldmath{$\Delta$}}}
\nc{\bDelta}{\mbox{\boldmath{$\Delta$}}}
\nc{\bdel}{\mbox{\boldmath{$\delta$}}}
\nc{\bdelta}{\mbox{\boldmath{$\delta$}}}
\nc{\bet}{\mbox{\boldmath{$\eta$}}}
\nc{\beps}{\mbox{\boldmath{$\epsilon$}}}
\nc{\bvep}{\mbox{\boldmath{$\vep$}}}
\nc{\bgam}{\mbox{\boldmath{$\gamma$}}}
\nc{\bgamma}{\mbox{\boldmath{$\gamma$}}}
\nc{\bGamma}{\mbox{\boldmath{$\Gamma$}}}
\nc{\bLam}{\mbox{\boldmath{$\Lambda$}}}
\nc{\bLambda}{\mbox{\boldmath{$\Lambda$}}}
\nc{\blambda}{\mbox{\boldmath{$\lambda$}}}
\nc{\bmu}{ \mbox{\boldmath{$\mu$}}}
\nc{\bOm}{ \mbox{\boldmath{$\Omega$}}}
\nc{\bOmega}{ \mbox{\boldmath{$\Omega$}}}
\nc{\bom}{ \mbox{\boldmath{$\omega$}}}
\nc{\bomega}{ \mbox{\boldmath{$\omega$}}}
\nc{\bpi}{ \mbox{\boldmath{$\pi$}}}
\nc{\bPi}{ \mbox{\boldmath{$\Pi$}}}
\nc{\bpsi}{ \mbox{\boldmath{$\psi$}}}
\nc{\bPsi}{ \mbox{\boldmath{$\Psi$}}}
\nc{\bphi}{ \mbox{\boldmath{$\phi$}}}
\nc{\bPhi}{ \mbox{\boldmath{$\Phi$}}}
\nc{\bxi}{ \mbox{\boldmath{$\xi$}}}
\nc{\bXi}{ \mbox{\boldmath{$\Xi$}}}
\nc{\bSig}{\mbox{\boldmath{$\Sigma$}}}
\nc{\bSigma}{\mbox{\boldmath{$\Sigma$}}}
\nc{\bsig}{\mbox{\boldmath{$\sigma$}}}
\nc{\btau}{\mbox{\boldmath{$\tau$}}}
\nc{\bThe}{\mbox{\boldmath{$\Theta$}}}
\nc{\bTheta}{\mbox{\boldmath{$\Theta$}}}
\nc{\bthe}{\mbox{\boldmath{$\theta$}}}
\nc{\btheta}{\mbox{\boldmath{$\theta$}}}
\nc{\bzeta}{\mbox{\boldmath{$\zeta$}}}
\nc{\bIm}{\mbox{\boldmath{$\Im$}}}

\nc{\ba}{ { \bf a }}
\nc{\bA}{ { \bf A }}
\nc{\bB}{ { \bf B }}
\nc{\bb}{ { \bf b }}
\nc{\bc}{ { \bf c }}
\nc{\bC}{ { \bf C }}
\nc{\bD}{ { \bf D }}
\nc{\bd}{ { \bf d }}
\nc{\be}{ { \bf e }}
\nc{\bF}{ { \bf F }}
\nc{\bG}{ { \bf G }}
\nc{\bh}{ { \bf h }}
\nc{\bH}{ { \bf H }}
\nc{\bI}{ { \bf I }}
\nc{\bJ}{ { \bf J }}
\nc{\bK}{ { \bf K }}
\nc{\bL}{ { \bf L }}
\nc{\bM}{ { \bf M }}
\nc{\bn}{ { \bf n }}
\nc{\bO}{ { \bf O }}
\nc{\bP}{ { \bf P }}
\nc{\br}{ { \bf r }}
\nc{\bR}{ { \bf R }}
\nc{\bs}{ { \bf s }}
\nc{\bS}{ { \bf S }}
\nc{\bT}{ { \bf T }}
\nc{\bt}{ { \bf t }}
\nc{\bu}{ { \bf u }}
\nc{\bU}{ { \bf U }}
\nc{\bv}{ { \bf v }}
\nc{\bV}{ { \bf V }}
\nc{\bW}{ { \bf W }}
\nc{\bw}{ { \bf w }}
\nc{\bx}{ { \bf x }}
\nc{\bX}{ { \bf X }}
\nc{\by}{ { \bf y }}
\nc{\bY}{ { \bf Y }}
\nc{\bz}{ { \bf z }}
\nc{\bZ}{ { \bf Z }}

\nc{\YR}{[\bY,R]}
\nc{\YgivenR}{[\bY \mid R]}
\nc{\RgivenY}{[R \mid \bY]}
\nc{\Y}{[\bY]}
\nc{\R}{[R]}

\nc{\dio}{d_i^o}
\nc{\timi}{t_{i,m_i}}
\nc{\betahat}{\hat{\bbet}}
\nc{\mui}{\bmu_{\rm I}}
\nc{\mue}{\bmu^{\rm E}}
\nc{\mup}{\bmu^{\rm P}}
\nc{\muihat}{\hat{\bmu}_{\rm I}}
\nc{\muehat}{\hat{\bmu}^{\rm E}}
\nc{\muphat}{\hat{\bmu}^{\rm P}}
\nc{\delhat}{\hat{\bdel}}
\nc{\muhat}{\hat{\bmu}}

\nc{\iid}{\stackrel{\rm iid}{\sim}}
\nc{\law}{\stackrel{\scl}{=}}

\nc{\phiij}{ \phi_{ij}( \Delta_0) }
\nc{\phiiprmj}{ \phi_{i'j}( \Delta_0) }
\nc{\phiijprm}{ \phi_{ij'}( \Delta_0) }
\nc{\phixy}{ \phi( X_i(S_{ik}), Y_j(T_{jl}) ) }
\nc{\phixydo}{ \phi( X_i(S_{ik}), Y_j(T_{jl})-\Delta_0 ) }
\nc{\phixyd}{ \phi( X_i(S_{ik}), Y_j(T_{jl})-\Delta) }
\nc{\phixydstar}{ \phi^*( X_i(S_{ik}), Y_j(T_{jl})-\Delta) }
\nc{\phixystdttil}{ \tilde{\phi}( X_i(s), Y_j(t)-\Delta, \theta) }
\nc{\phixydttil}{ \tilde{\phi}( X_i(S_{ik}), Y_j(T_{jl})-\Delta, \theta) }
\nc{\Nmn}{{\sqrt{N} \over mn}}

\nc{\Xis}{X_i(s)}
\nc{\Yjt}{Y_j(t)}

\nc{\bthehat}{\hat{\bthe}}

\nc{\Ritil}{\tilde{R}_i}

\nc{\Ybar}{\overline{Y}}
\nc{\Rbar}{\overline{R}}
\nc{\Nbar}{\overline{N}}
\nc{\intzeroinf}{\int_0^\infty}

\nc{\Fhat}{\hat{F}}
\nc{\Ghat}{\hat{G}}

\nc{\FhatS}{\hat{F}(S_{ik})}
\nc{\GhatT}{\hat{G}(T_{jl})}

\nc{\Fhatik}{\hat{F}_{ik}}
\nc{\Ghatjl}{\hat{G}_{jl}}
\nc{\Fik}{F_{ik}}
\nc{\Gjl}{G_{jl}}
\nc{\phiijkl}{\phi_{ik,jl}(\Delta)}
\nc{\phiijkltil}{\tilde{\phi}_{ik,jl}(\Delta_0,\theta_0)}
\nc{\ord}{N^{-3/2}}           
\nc{\sumijkl}{\sum_{ijkl}}

\nc{\Citil}{\tilde{C}_i}
\nc{\Crtil}{\tilde{C}_r}
\nc{\Djtil}{\tilde{D}_j}
\nc{\Ditil}{\tilde{D}_i}

\nc{\Cithe}{\tilde{C}^{\theta}_i}
\nc{\Djthe}{\tilde{D}^{\theta}_j}

\nc{\Sikthe}{S_{ik}^{\theta}}
\nc{\Tjlthe}{T_{jl}^{\theta}}

\nc{\Zi}{ \bZ_{-i}}
\nc{\zic}{ \lbc z(\bs_j) \: : \: i \neq j \rbc }
\nc{\zkap}{ \bz_{\kappa} }
\nc{\sumi}{ \sum_i }
\nc{\sumj}{ \sum_j }
\nc{\sumij}{ \sum_{i < j} }
\nc{\sumiandj}{ \sum_{i, j} }
\nc{\zsi}{ z(\bs_i) }
\nc{\Zsi}{ Z(\bs_i) }
\nc{\zsj}{ z(\bs_j) }
\nc{\zsn}{ z(\bs_n) }
\nc{\zsone}{ z(\bs_1) }
\nc{\pZ}{ \Pr \lbc \bZ \rbc }
\nc{\qz}{ Q( \bz ) }
\nc{\qZ}{ Q( \bZ ) }

\nc{\thetaYD}{\theta_{Y\mid D}}
\nc{\thetaD}{\theta_D}
\nc{\psiDY}{\psi_{D\mid Y}}
\nc{\psiY}{\psi_Y}

\nc{\tn}{\Theta^{\nu}}
\nc{\Etn}{E_{\theta^{\nu}}}
\nc{\tnone}{\Theta^{\nu+1}}
\nc{\Lm}{L_{\text{m}}}
\nc{\Lo}{L_{\text{o}}}
\nc{\Ym}{Y_{\text{m}}}
\nc{\Yo}{Y_{\text{o}}}
\nc{\ym}{y_{\text{m}}}
\nc{\yo}{y_{\text{o}}}

\nc{\vijb}{v_{ij} - \bX_{i(j)}  \bbet}
\nc{\vikb}{v_{ik} - \bX_{i(k)}  \bbet}
\nc{\vilb}{v_{il} - \bX_{i(l)}  \bbet}
\nc{\betart}{ \bbet^{(r)}_{t_i} }
\nc{\betarj}{ \bbet^{(r)}_j }
\nc{\yij}{y_{ij}}
\nc{\Xmisi}{ {\bX_{ i{\rm (mis)} }} }
\nc{\Xobsi}{ {\bX_{ i{\rm (obs)} }} }
\nc{\Zobsi}{ {\bZ_{ i{\rm (obs)} }} }
\nc{\bSigobs}{ \bSig_{  {\rm obs} } }
\nc{\bSigmis}{ \bSig_{  {\rm mis} } }
\nc{\bSigmo}{ \bSig_{  {\rm mis,obs} } }
\nc{\bSigom}{ \bSig_{  {\rm obs,mis} } }
\nc{\Xil}{{\bX}_{il}}
\nc{\Zil}{{\bZ}_{il} }
\nc{\omilr}{\omega_{il}^{(r)}}
\nc{\delio}{\bdel_i^{{\rm obs}} }

\nc{\yio}{ {y_i^{\rm o} }}
\nc{\Yio}{ {Y_i^{\rm o }} }
\nc{\Yim}{ {Y_i^{\rm m} }}
\nc{\yim}{ {y_i^{\rm m} }}
\nc{\Yc}{Y^{\rm c}}
\nc{\Yic}{Y_i^{\rm c}}
\nc{\yc}{y^{\rm c}}
\nc{\yic}{y_i^{\rm c}}

\nc{\yi}{y_i}
\nc{\Yi}{Y_i}

\nc{\fyic}{f ( \yic ; \; \psiY )}
\nc{\fyi}{f ( y_i ; \; \psiY ) }
\nc{\fdigivenyic}{f ( d_i  \mid  \yic ; \; \psiDY )}
\nc{\fditilgivenyic}{f ( \tilde{d}_i  \mid  \yic ; \; \psiDY )}
\nc{\fditilgivenyi}{f ( \tilde{d}_i  \mid  \yi ; \; \psiDY )}
\nc{\Fditilgivenyic}{F ( \tilde{d}_i  \mid  \yic ; \; \psiDY )}
\nc{\Fditilgivenyi}{F ( \tilde{d}_i  \mid  \yi ; \; \psiDY )}
\nc{\fdigivenyi}{f (d_i \mid y_i ; \;  \psiDY  )}
\nc{\fyicdi}{f \left( \yic, d_i \right)}
\nc{\fyidi}{f \left( \yi, d_i \right)}

\nc{\fymidr}{f_{Y \mid R}}
\nc{\fyr}{f_{Y,R}}
\nc{\frmidy}{f_{R \mid Y}}
\nc{\fy}{f_Y}
\nc{\fr}{f_R}

\nc{\fyicgivendi}{f (\yic \mid d_i; \; \thetaYD )}
\nc{\fyigivendi}{f (\yi \mid d_i; \; \thetaYD )}
\nc{\fyicgivens}{f (\yic \mid s; \; \thetaYD )}
\nc{\fyigivens}{f (\yi \mid s; \; \thetaYD )}
\nc{\fdi}{f ( d_i; \; \thetaD )}

\nc{\fyicX}{f ( \yic \mid X_i; \; \psiY )}
\nc{\fyiX}{f ( y_i \mid X_i; \; \psiY ) }
\nc{\fdigivenyicX}{f ( d_i  \mid  \yic, X_i ; \; \psiDY )}
\nc{\fdigivenyiX}{f (d_i \mid y_i, X_i ; \;  \psiDY  )}
\nc{\fyicdiX}{f \left( \yic, d_i \mid X_i \right)}

\nc{\fyicgivendiX}{f (\yic \mid d_i, X_i; \; \thetaYD )}
\nc{\fyigivendiX}{f (y_i \mid d_i, X_i; \; \thetaYD )}
\nc{\fdiX}{f ( d_i \mid X_i; \; \thetaD )}

\nc{\Yistar}{\bY_i^*}

\nc{\Dio}{D_i^{\rm obs}}
\nc{\bdelio}{\bdel_{ i \, {\rm (obs)}} }

\nc{\fygivend}{f_{Y \mid \delta}}
\nc{\fyd}{f_{Y, \delta}}
\nc{\fd}{f_\delta}
\nc{\FD}{F_D}
\nc{\fygivenbd}{f_{Y\mid b, \delta}}

\nc{\alphahat}{\hat{\bal}}
\nc{\phihat}{\hat{\bphi}}
\nc{\thetahat}{\hat{\bthe}}
\nc{\thetatilde}{\tilde{\bthe}}
\nc{\scoretheta}{\bS(\bthe; \, \scc)}
\nc{\hesstheta}{\bH(\bthe; \, \scc)}
\nc{\infotheta}{\sci(\bthe; \, \scc)}
\nc{\sitheta}{\bs_i(\bthe; \, \scc_i)}
\nc{\sithetahat}{\bs_i(\thetahat; \, \scc_i)}

\nc{\loglikobs}{\ell_{{\rm o}}(\bthe; \, \sco)}
\nc{\scoreobs}{\bS_{{\rm o}}(\bthe; \, \sco)}
\nc{\hessobs}{\bH_{{\rm o}}(\bthe; \, \sco)}
\nc{\infoobs}{\scj_{{\rm o}}(\bthe; \, \sco)}

\nc{\Cil}{\scc_{il}}
\nc{\olog}{\lambda^*(\bthe, \Xobs)}
\nc{\LthetaC}{\scl(\bthe; \, \scc)}
\nc{\LthetaCi}{\scl_i(\bthe; \, \scc_i)}
\nc{\LthetaCil}{\scl_i (\bthe; \, \scc_{il}) }
\nc{\lthetaC}{\ell(\bthe; \, \scc)}
\nc{\lthetaCi}{\ell_i(\bthe; \, \scc_i)}
\nc{\lthetaCil}{\ell_i (\bthe; \, \scc_{il}) }
\nc{\Qtheta}{\scq \left( \bthe \, \left| \,  \bthe^{(r)} \right. \right)}
\nc{\thetar}{\bthe^{(r)}}
\nc{\thetas}{\bthe^{(s)}}
\nc{\alphas}{\bal^{(s)}}
\nc{\psis}{\psi^{(s)}}
\nc{\alphasplusone}{\bal^{(s+1)}}
\nc{\psisplusone}{\bpsi^{(s+1)}}
\nc{\alphapsis}{\left( \alphas, \psis \right)}

\nc{\thetarplusone}{\bthe^{(r+1)}}
\nc{\ologi}{\lambda^*_i(\bthe, \Xobs)}
\nc{\llogi}{\lambda_i \left( \bthe, \tilde{\Xcom}_{il} \right) }

\nc{\scxil}{\tilde{\Xcom}_{il}}
\nc{\siginv}{\bSig_i^{-1}}

\nc{\fofym}{ f \left( \by_i \mid \bbet_m, \bSig \right) }
\nc{\mphim}{ \phi_M \lsq \bSig^{-1/2}(\by_i - \bX_i \bbet_m) \rsq }
\nc{\mphit}{ \phi_M \lsq \bSig^{-1/2}(\by_i - \bX_i \bbet_{t_i}) \rsq }
\nc{\mphij}{ \phi_M \lsq \bSig^{-1/2}(\by_i - \bX_i \bbet_j) \rsq }
\nc{\mphik}{ \phi_M \lsq \bSig^{-1/2}(\by_i - \bX_i \bbet_k) \rsq }
\nc{\expkerm}{ \exp  \lbc -\half \bu_i(\bbet_m)' \bSig^{-1} \bu_i(\bbet_m)
  \rbc }
\nc{\expkerk}{ \exp  \lbc -\half \bu_i(\bbet_k)' \bSig^{-1} \bu_i(\bbet_k)
  \rbc }
\nc{\expkerj}{ \exp  \lbc -\half \bu_i(\bbet_j)' \bSig^{-1} \bu_i(\bbet_j)
  \rbc }
\nc{\normscorem}{\left( \bX_i' \bSig^{-1} \bX_i \bbet_m - \bX_i' \bSig^{-1}
  \by_i \right) }
\nc{\normscorej}{\left( \bX_i' \bSig^{-1} \bX_i \bbet_j - \bX_i' \bSig^{-1}
  \by_i \right) }
\nc{\piti}{ \pi \left( t_i, \bal, \bZ_i\bgam \right) }
\nc{\omij}{ \om_{ij} \left( t_i, \bal, \bZ_i\bgam \right) }
\nc{\phibetak}{ \phi_M(\bbet_k) }
\nc{\phibetaj}{ \phi_M(\bbet_j) }
\nc{\dphidbetak}{ \left. \dpl \phibetak \right/ \dpl \bbet_k }
\nc{\dphidbetakf}{ \frac{ \dpl \phibetak }{ \dpl \bbet_k } }
\nc{\uik}{\bu_i \left( \bbet_k  \right)}
\nc{\mset}{ \{ 0, 1, \ldots, M \} }
\nc{\betasigma}{ \left( \lbc \bbet^{(r)}_t \rbc, \bSig^{(r)} \right) }
\nc{\Thetar}{ \bThe^{(r)} }

\nc{\shatkm}{\hat{S}_{\rm KM}}

\nc{\ds}{\displaystyle}

\nc{\beq}{\begin{eqnarray*}}
\nc{\eeq}{\end{eqnarray*}}

\nc{\beqna}{\begin{eqnarray}}
\nc{\eeqna}{\end{eqnarray}}

\nc{\bct}{\begin{center}}
\nc{\ect}{\end{center}}

\nc{\bds}{\begin{description}}
\nc{\eds}{\end{description}}

\nc{\bit}{\begin{itemize}}
\nc{\eit}{\end{itemize}}

\nc{\bnu}{\begin{enumerate}}
\nc{\enu}{\end{enumerate}}

\nc{\bgt}{\begin{table}}
\nc{\bgtb}{\begin{center} \begin{tabular}}
\nc{\entb}{\end{tabular} \end{center} }
\nc{\ent}{\end{table}}

\nc{\ts}{\textstyle}

\nc{\bgl}{\begin{letter}}
\nc{\op}{\opening}
\nc{\incl}{\input{sendout.ltr} \closing{Best Regards,} \end{letter} }

\title{Modelling  disease progression  with multi-level  electronic health records data and informative observation times: an application to treating iron deficiency anaemia in primary care of the UK }

\author{\textit{Li Su}\thanks{Correspondence: \textit{li.su@mrc-bsu.cam.ac.uk}}~\thanks{MRC Biostatistics Unit, School of Clinical Medicine, University of Cambridge,  Robinson Way, Cambridge CB2 0SR, UK}, \textit{Yafeng Cheng}\footnotemark[2],  \textit{Dora I.A. Pereira}\thanks{Department of Pathology, University of Cambridge, Tennis Court Road, Cambridge, CB2 1QP, UK}  and \textit{Jonathan J. Powell}\thanks{Department of Veterinary Medicine, University of Cambridge, Madingley Road, Cambridge, CB3 0ES, UK}}

\date{}

\begin{document}
\maketitle

\begin{center}
\noindent {\bf Abstract}
\end{center}

Modelling  disease progression of iron deficiency anaemia (IDA) following oral iron supplement prescriptions is a prerequisite for evaluating the cost-effectiveness of oral iron supplements. Electronic health records (EHRs) from the Clinical Practice Research Datalink (CPRD) provide rich longitudinal data on IDA disease progression in  patients registered with 663 General Practitioner  (GP)  practices in the UK, but  they also create challenges in statistical  analyses. First, the CPRD data are clustered at multi-levels (i.e., GP practices and patients), but their large volume makes it computationally difficult to implement estimation of standard  random effects models for multi-level data. Second, observation times in the CPRD data are irregular and could be informative about the disease progression. For example, shorter/longer gap times between GP visits could be associated with deteriorating/improving IDA. Existing methods to address informative observation times are mostly based on complex joint models, which adds more computational burden. To tackle these challenges, we develop a computationally efficient approach to modelling disease progression with EHRs data while accounting for variability at multi-level clusters and informative observation times. We apply the proposed method to the CPRD data to investigate IDA improvement and treatment intolerance following oral iron prescriptions in primary care of the UK.

{\emph{\textbf{Keywords}: big data; bootstrap; composite likelihood; irregular longitudinal data; multi-level data; outcome-dependent follow-up.}

\baselineskip=24pt
\section{Introduction}\label{Intro}

Iron deficiency affects 24\% of adolescent girls and 12\% of pre-menopausal women in the UK, and anaemia prevalence in these populations is nearly 10\% \cite[]{NDNS}. Oral iron is generally the first-line treatment for iron deficiency anaemia (IDA). However, currently prescribed oral iron supplements in the UK can be poorly tolerated  \cite[]{Tolkien2015}, which affects their adherence \cite[]{Saha2007,Lindgren2009,Souza2009,Zaim2011,Tolkien2015} and can generate further  costs to the National Health Service (NHS)  due to associated adverse effects (e.g., gastrointestinal symptoms). Therefore, it is of marked interests to  evaluate the cost-effectiveness of existing oral iron supplements and inform  decisions regarding developments of  new oral irons with reduced adverse effects or switching to alternative intravenous strategies. As a prerequisite, it is necessary to  investigate IDA disease progression (i.e., IDA improvement and  treatment intolerance) following oral iron prescriptions, ideally using primary care data of the UK, e.g., from the Clinical Practice Research Datalink (CPRD).

The CPRD provides longitudinal electronic health records (EHRs) from patients registered with 663 General Practitioner  (GP)  practices in the UK, which are invaluable  to address questions on IDA disease progression. However, there are two challenges in analyzing the CPRD data for IDA. First, the CPRD data are clustered at multi-levels (i.e., GP practices and patients), but their large volume  makes it computationally difficult to implement  estimation of standard  random effects models for multi-level data \cite[]{Rabe-Hesketh2006}. In our exploratory analyses of the CPRD data for IDA,  off-the-shelf statistical software cannot achieve the  computation  for non-linear random effect models at large scale (323,718 observations from 120,892 patients) in a timely fashion (see details in Section~\ref{CPRDdata}). As a result, it is necessary to develop a computationally efficient approach to analyzing the CPRD data for IDA while accounting for variability at multi-level clusters.  

Second, observation times in the CPRD data are irregular and could be informative for IDA disease progression.   Specifically, because the CPRD data were  collected when patients visited GP practices,  the follow-up lengths and visiting times varied considerably across patients. When visits were initiated by patients due to IDA symptoms,  shorter gap times between GP visits could be associated with deteriorating IDA. Similarly, patients with improved haemoglobin levels after oral iron prescriptions might be less motivated to come back to GP practices for a haemoglobin blood test, which leads to  the association between longer gap times  and IDA improvement.  In Section~\ref{CPRDdata}, we provide more details about this phenomenon in the IDA context. In general, selection bias from irregular and possibly informative observation times is a common problem in routinely collected data such as EHRs data or data from clinic-based studies \cite[]{Pull2016}. In order to provide valid inference about  longitudinal outcome processes  (e.g., IDA disease progression in our case) using routinely collected data, it is necessary to handle  informative observation times.  However, existing methods to address informative observation times are mostly based on complex joint models \cite[]{Sun2007}, which adds more computational burden to the analyses of the CPRD data. In Section~\ref{related}, we provide a brief review of existing methods on this topic.

Motivated by these challenges from  the CPRD data for IDA, in this paper we develop a computationally efficient approach to modelling disease progression with   multi-level EHRs data and informative observation times.  Specifically, we propose a discrete-time Markov model with multinomial logistic regressions to estimate transition probabilities between  disease states that characterize the disease progression process. To account for informative observation times and other features of the visiting process as in the IDA example, we use regressions to directly adjust for variables such as the treatment sequence within patients, the time in follow-up during a course of treatment and the gap time between assessment visits within a course of treatment. This avoids building complex joint models for the disease progression and the visiting processes, which are difficult to implement given the large volume of EHRs data.  For computational  efficiency, point estimation of the model parameters is based on composite likelihood without random effects \cite[]{Varin2011}, while  bootstrap-based methods,  including the computationally efficient `estimating function bootstrap' (EFB) \cite[]{Hu2000,Roberts2009a}, are adapted to provide confidence intervals that take into account  variability at multi-level clusters. Simulations in Section~\ref{simulation} show that the EFB performs equally well as the  nonparametric bootstrap in settings similar to the CPRD data.

The rest of the paper is organized as follows. In Section~\ref{CPRDdata}, we introduce the CPRD data and provide details of the challenges arising from them. A review of related methods for irregular longitudinal data and informative observation times is provided in Section~\ref{related}.
In Section~\ref{methods} we describe the proposed methodology.  In Section~\ref{analysis}, we apply our methods to the CPRD data and examine the associations of  covariates with IDA improvement and treatment intolerance  in primary care of the UK.   A simulation study for evaluating the performance of the bootstrap methods used in the CPRD analysis is presented in Section~\ref{simulation}.
Finally, we conclude with a discussion in Section~\ref{conclusion}.

\section{The CPRD data}\label{CPRDdata}

The CPRD is a governmental, not-for-profit research service in the UK to provide anonymised primary care data for public health research. Established in 1987 and currently representing almost $10\%$ of the UK population,  it is one of the largest databases of longitudinal medical records from primary care in the world \cite[]{Herrett2015}. The CPRD provides the computerised medical records maintained by GPs in the UK. GPs play a key role in the NHS, as they are responsible for primary health care and specialist referrals. Thus data recorded in the CPRD include demographic information, prescription details, clinical events, preventative care provided, specialist referrals, hospital admissions and  major outcomes.  This collection of clinical information is  further complemented through secure anonymised linkage to secondary care databases, such as hospital events from the Hospital Episode Statistics (HES).

In this paper, we focus on  longitudinal data extracted from the CPRD for pre-menopausal women  prescribed with  ferrous iron salts between January 2000 and October 2014. These CPRD data are further linked with the hospital records from the HES. In total, our study population consists of 120,892 women aged 18-45 years (i.e., typically pre-menopausal) who had at least one prescription of  one ferrous iron salt (i.e., sulphate, fumarate and gluconate) and at least one follow-up  assessment since the prescription from 663 GP practices during the study period (2000-2014). 
The aim is  to investigate IDA improvement and treatment intolerance (e.g.,  hospital referral) following the prescription of an iron salt,  and to examine associated covariate factors (e.g., iron salt type and iron dose) . 

Specifically, four disease states based on information available from the CPRD and HES are defined: 
no improvement of haemoglobin (State 1);
 improvement of haemoglobin (State 2);
 hospital referral (State 3);
 anaemia resolved (State 4).
Table~\ref{table1} presents the observed transition matrix for the four disease states at assessment visits within  the courses of treatment, aggregated over all patients. A course of treatment is defined by a new prescription of oral iron distinct from the immediate previous oral iron  prescription (by type or by iron dose) or a prescription of the same oral iron as before but with the new prescription date more than 2 months from last prescription date. Here we summarize the transitions within the follow-up of a specific course of treatment  because some patients had  long gaps between two courses of treatment  possibly due to relapse after a long period of  time since improvement of  haemoglobin  in their previous courses of treatment. Note that State 4 is an absorbing state and no follow-up continued after entering this state. 

 \begin{table}[ht]
\caption{Observed transition matrix for the IDA disease states  in the CPRD data. State 1: no improvement of haemoglobin; State 2: 
 improvement of haemoglobin;
 State 3: hospital referral;
 State 4: anaemia resolved. }
\centering
\begin{tabular}{rrrrr}
  \hline
     & to State &&&\\
 from State & 1 & 2 & 3 & 4 \\ 
  \hline
1 & 122624 & 62905 & 36184 & 32623 \\ 
 2 & 6350 & 21272 & 5614 & 5826 \\ 
  3 & 7059 & 6434 & 14379 & 2448 \\ 
   \hline
\end{tabular}\label{table1}
\end{table}

\bigskip

\begin{table}[hb]
\caption{Summary of covariate distributions in the CPRD data }
\centering
{\small \begin{tabular}{lll}
  \hline
  Number of patients && 120,892\\
  Number of courses of treatment & &196,654\\
  Compound class of iron salts (\%) &&  \\
  &Class 1: ferrous fumerate & 83671 (42.5)\\
  &Class 2: ferrous gluconate  & 9451 (4.8)\\
  &Class 3: ferrous sulphate & 97726 (49.7)\\
  &Class 4: ferrous sulphate modified-release & 5806 (3)\\
  IMD2010 of the CPRD patients (\%) & &   \\
   &1st quintile:  least deprived  & 14019 (11.6)\\
   &2nd quintile &  14025 (11.6)\\
   &3rd quintile   & 14377 (11.9)\\
   &4th quintile & 16540 (13.7)\\
  &5th quintile: most deprived   & 16519  (13.7)\\
   &NA & 45412 (37.6)\\
  \hline
\end{tabular}}\label{table2}
\end{table}

The covariate factors of interest  for anaemia improvement and treatment intolerance include compound classes of iron salts, daily iron doses in the iron salts and socio-economic deprivation level of the region where the patient resided, measured by the quintiles of the 2010 English Index of Multiple Deprivation for England (IMD2010: 1=least deprived, ..., 5=most deprived).  
Table~\ref{table2} presents the percentages  of  iron salt compound classes among the total 196,654 courses of treatment and  the distribution of IMD2010 deprivation index for the 120,892 patients.  The oral ferrous iron prescribed in primary care of the UK could be categorized into 4 classes, namely, ferrous fumerate (Class 1),  ferrous gluconate (Class 2), ferrous sulphate (Class 3) and ferrous sulphate modified-release (Class 4). As such, the vast majority of the oral iron prescriptions in the UK are either  ferrous fumarate or ferrous sulphate  (Table 2).  In the analysis presented in 
Section~\ref{analysis}, we will focus on the comparison between ferrous fumarate (Class 1) and  ferrous sulphate (Class 3), due to small sample sizes and different ranges of iron dose for ferrous gluconate (Class 2) and ferrous sulphate modified-release (Class 4) (see Figure~1 in the Supplementary Materials). We provide further details about  covariates in the CPRD data in Section~1 of the Supplementary Materials.

In total, our data contain 323,718 assessments from 120,892 patients at 663 GP practices who had at least one follow-up assessment (either for a haemoglobin blood test or hospital referral) since the course of treatment started (i.e., when oral iron  was prescribed) and whose disease states can be ascertained at these assessments. As mentioned in Section~\ref{Intro}, it is difficult to fit standard random effect models to multi-level data at   such large scale in a timely manner. In addition, the CPRD data are highly unbalanced with differential follow-up lengths and visiting times across patients.   
The number of patients by GP practices is highly varied, as shown in Figure~\ref{figure2} (a). Within patients, there are differential numbers of courses of treatment during  the study  period. Figure~\ref{figure2} (b) shows that the number of courses of treatment within patients is highly varied as well, with the majority of patients having fewer than 4 courses of treatment in the study period. Moreover, there are a lot of variations in the lengths of follow-up  and the gap times between consecutive assessments within courses of treatment, as shown in Figure~\ref{figure2} (c) and (d), respectively. 
\begin{figure}[htp]
\centering
\centering\includegraphics[scale=0.7]{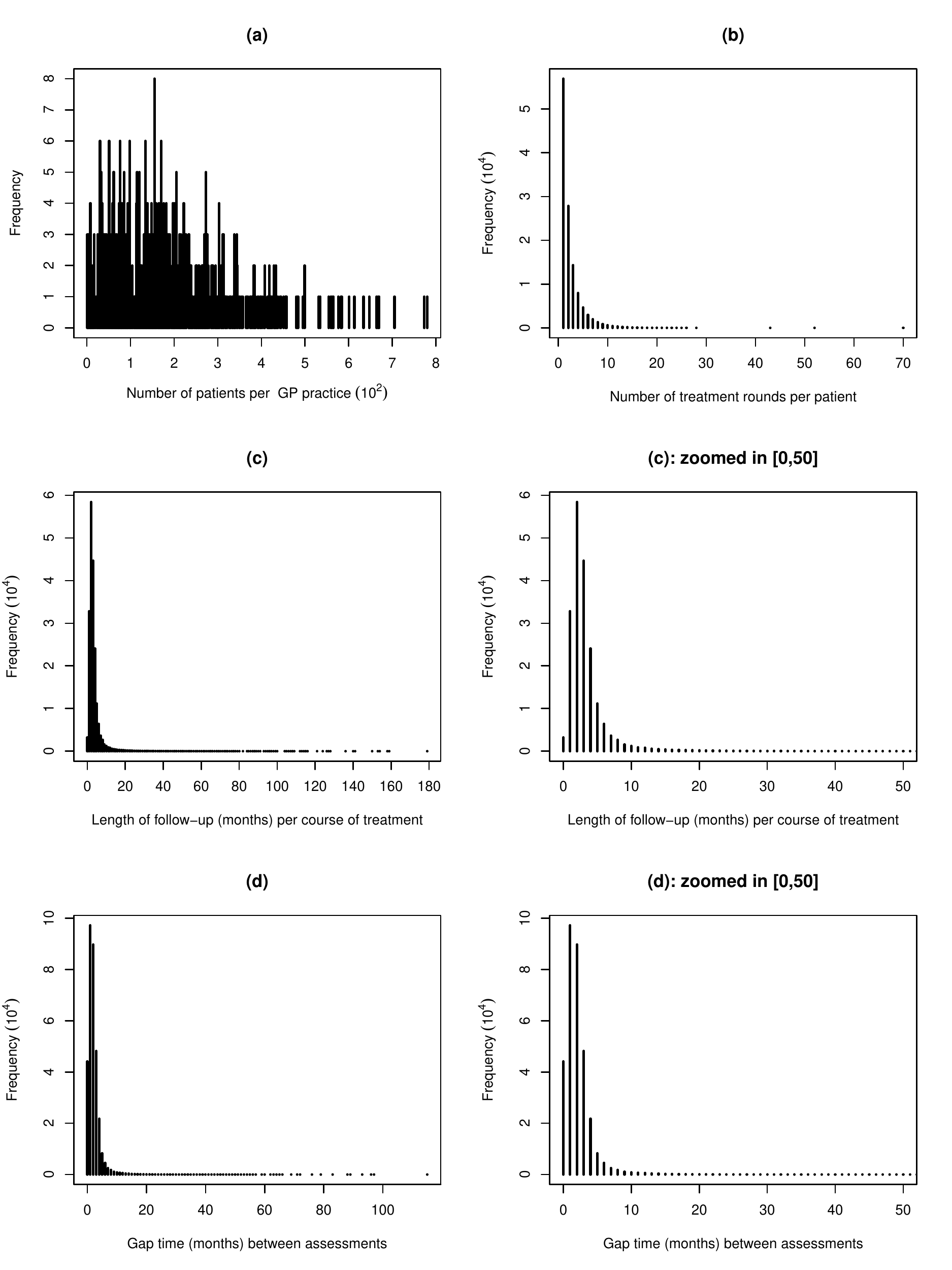}
\caption{Observation patterns in the CPRD data: (a) number of patients per  GP practice; (b) number of courses of treatment  per  individual patient; (c) length of follow-up (in months) per course of treatment; (d) gap times (in months) between consecutive assessments within courses of treatment. }\label{figure2}
\end{figure}

In addition to the concern about  informative observation times discussed in Section~\ref{Intro}, other features of the visiting process in the CPRD data, such as the number of courses of treatment and follow-up lengths within courses of treatment, could also be associated with the IDA disease progression process. Specifically, some  patients were prescribed with different classes of oral iron primarily due to ineffectiveness of or intolerance to previous oral iron class. As a result,  disease progression (i.e.,  IDA improvement and treatment intolerance) at later courses of treatment could be quite different
from that at earlier courses of treatment. Similarly, disease progression within a specific course of treatment can be time-dependent. We will address these issues in the proposed approach presented in Section~\ref{methods}.

\section{Related work: methods for irregular longitudinal data}\label{related}

Before introducing our proposed approach in Section~\ref{methods}, we briefly review the literature for  irregular longitudinal data, which is relevant to the  `informative observation times' problem in the CPRD data.
 
In long-term follow-up studies, irregularly measured longitudinal data can result from outcome-dependent visiting times, where patients with a history of poor health outcomes are being assessed with greater frequency and regularity.
There is  an extensive literature on irregular longitudinal data and outcome-dependent visiting times;  see \cite{Pull2016} for a recent comprehensive review. In general, there are three types of approach. First,  standard approaches include   generalized estimating equations (GEE), mixed models, and Lin-Ying's estimating equations \cite[]{Lin2001}, where   consistent estimates of regression coefficients in the model for longitudinal outcome can  be obtained under certain ignorability assumptions about the visiting process. Second, weighting longitudinal observations by inverse intensity of the visiting process can accommodate the scenarios where the visiting process only depends on observed information (including previous outcomes) immediately up to the visiting time \cite[]{Lin2004}. Third, semiparametric and parametric joint models for the outcome and  visiting processes have also been proposed, where shared or correlated random effects are used to characterize the associations between the two processes \cite[]{Sun2007}. These joint models allow the visiting process to depend on the \textit{current} longitudinal outcome after conditioning on  observed information up to the current time. Therefore they are able to handle informative observation times such as in the CPRD data. In the context of modelling disease progression, \cite{Lange2014} proposed a joint model for the  disease progression and  visiting processes, where the disease progression process follows a continuous-time multi-state model and the visiting process is a Markov-modulated Poisson process with the rates directly depending on the latent disease states.

In this paper, we point out that for modelling disease progression, it is  natural to 
incorporate observed gap times between assessment visits as an  adjustment variable into the model for disease progression. This is different from the scenarios covered in most of the existing methods, where the marginal mean of the longitudinal outcome is of interest \cite[]{Pull2016}. For example,  the gap time can enter the regression structure of a discrete-time Markov  model such that the  probabilities of transitioning to different disease states directly depend on the time elapsed since last visit. In this case,  \textit{current} disease states are allowed to depend on the visiting process.
  In fact, directly conditioning on the gap time while modelling disease progression is very similar to the \textit{pattern mixture modelling} approach to dealing with informative missing data, where conditional distributions of the longitudinal outcome  given the  missing data patterns are modelled but no model for the missing data mechanism is required \cite[]{Daniels2008}. Therefore, this `conditioning' approach does not require specifying a model for the visiting process (i.e., the observation times). Moreover, computationally it is an efficient approach   in  our setting,  given the large volume of the CRPD data and the computational burden from implementing existing methods for informative observation  times (e.g., fitting joint models with random effects).  One limitation of this approach is that the interpretations of the covariate effects  have to be conditional on the adjustment variables for the visiting process, but this is not of primary concern in the  IDA context.

\section{Methodology}\label{methods}
In this section, we describe the proposed approach to modelling disease progression in the IDA context.  Specifically, we build a  discrete-time Markov model with multinomial logistic regressions to  estimate  transition probabilities of  IDA disease states and their associations with covariates, while adjusting for variables that characterize the visiting process. The point estimation is based on composite likelihood,  where all observed transitions across the GP practices and  patients are assumed to be independent \cite[]{Varin2011}.  The inference is obtained through bootstrappoing, which  is performed at the highest level of the clusters (e.g., by GP practice in the IDA example) to preserve the correlation structure of the observations within the clusters. Because the direct non-parametric bootstrap with replacement requires refitting the model for each bootstrap sample, we also  adapt the computationally efficient estimating function bootstrap \cite[]{Hu2000,Roberts2009a} to our setting.

\subsection{Notation and model}
Let $S_{gijk}=s$ ($s\in \mathcal{S}=\{1,2,3,4\}$ ) denote the disease state at the visiting time $t_{gijk}$ in the $j$th course of treatment of the $i$th patient with IDA at the $g$th GP practice, where  $g=1, \ldots, G$, $i=1, \ldots, n_g$, $j=1,\ldots, n_{gi}$ and $k=1,\ldots, m_{gij}$. Recall that $s=1, 2, 3, 4$ represent, respectively,  no improvement in haemoglobin, 
improvement in haemoglobin,
 hospital referral and
 anaemia resolution.
The visiting time $t_{gijk}$ is irregularly spaced and we define the gap time between visits as $v_{gijk}=t_{gij,k+1}-t_{gijk}$. Let $\bX_{gij}$ denote the  covariate vector evaluated at the start of the course of treatment, for example, the compound class and daily dose of oral iron  prescribed, and the deprivation index in the region where the patient lived.  $\bX_{gij}$ is time-invariant during the follow-up  period of the corresponding course of treatment.

We assume that the disease progression process within each course of treatment for a patient follows a discrete-time Markov model. In addition, the censoring of patients' follow-up is assumed to be non-informative due to reasons such as patients moved out of the GP practice catchment areas or reaching the study end in December 2014. Specifically, 
we assume that the transition probability  from the disease state at $t_{gijk}$ to the state at $t_{gij,k+1}$ follows a multinomial logistic model 
\begin{eqnarray}\label{model}
&&\log\left\{\frac{\mbox{Pr}\left(S_{gij,k+1}=s' \mid S_{gijk}=s, \bZ_{gijk}\right)}{\mbox{Pr}\left(S_{gij,k+1}=1 \mid S_{gijk}=s,\bZ_{gijk}\right)}\right\} \nn\\
&=&\beta_{0}^{ss'}+\beta_{1}^{ss'} f_j(j)+\beta_{2}^{ss'} f_t(t_{gijk})+ \beta_{3}^{ss'} f_v(v_{gijk})+\bX_{gij}\trans\balpha^{ss'},
\end{eqnarray}
where $s\in \{1,2,3\}$, $s'\in \{2,3,4\}$,  $f_j(\cdot)$, $f_t(\cdot)$ and $f_v(\cdot)$ are functions of the index of the course of treatment $j$, the current time $t_{gijk}$ and the gap time to next visit $v_{gijk}$, respectively,  and  $\bZ_{gijk}=(j, t_{gijk}, v_{gijk}, \bX_{gij}\trans)\trans$.  Here  the dependence on the sequence of courses of treatment $j$ is analogous to the scenario of recurrent events in a survival analysis, where the hazards for recurrent events can vary by how many events  have occurred previously \cite[]{Cook2007}. The dependence on the time in the follow-up of a course of treatment is similar to allowing the hazard function to be  time-dependent in a survival analysis.

 All intercept terms and regression coefficients $\beta_{0}^{ss'}$, $\beta_{1}^{ss'}$, $\beta_{2}^{ss'}$, $\beta_{3}^{ss'}$ and 
$\balpha^{ss'}$ are allowed to vary by the transitions between  states $s$ and $s'$.  As discussed, we do not incorporate random effects at  the patient level and GP practice level. The covariate effects of interest $\balpha^{ss'}$  represent the population-averaged associations between $\bX_{gij}$ and the transition probabilities from   $S_{gijk}=s$ to $S_{gij,k+1}=s'$ after adjusting for variables that characterizing the visiting process. 
In Section~\ref{analysis} we will discuss the choices of $f_j(\cdot)$, $f_t(\cdot)$ and $f_v(\cdot)$ for the adjustment variables  in the IDA context.

\subsection{Point estimation}
For point estimation of the model parameters, we adopt the composite likelihood approach without considering  the correlations between transitions of disease states within patients and within GP practices \cite[]{Varin2011}. Specifically, the likelihood contribution of the $i$th patient in the $g$th GP practice is
\begin{equation}\label{lik_prod}
 \mathcal{L}_{gi}\left(\btheta\right)=\prod_{j=1}^{n_{gi}}\prod_{k=1}^{m_{gij}-1}\mbox{Pr}\left(S_{gij,k+1}\mid S_{gijk},\bZ_{gijk};\btheta \right), 
\end{equation} where   $\btheta$ denotes all parameters specified in the multinomial logistic model in~\eqref{model}. The total composite likelihood for maximization is 
 \begin{equation}\label{lik_prod2}
 \mathcal{L}\left(\btheta\right)=\prod_{g=1}^{G}\prod_{i=1}^{n_g}\mathcal{L}_{gi}\left(\btheta\right). 
\end{equation} Because the multinomial logistic model in~\eqref{model} is conditional on the current disease state $S_{gijk}$ and all parameters in the multinomial logistic model in~\eqref{model} are distinct by specific transitions,  the composite likelihood related to the transitions from $S_{gijk}\in \{1, 2, 3\}$ can be maximized separately to obtain the corresponding parameter estimates in~\eqref{model}. 
We use the  \texttt{nnet} package  \cite[]{Venables2002} in \texttt{R} for point estimation in the CPRD data analysis reported in Section~\ref{analysis}.

\subsection{Confidence intervals}
Confidence intervals for maximum composite likelihood estimates $\hat{\btheta}$ need to take into account 
clustering at the patient and GP practice levels. In this paper we use bootstrap methods to construct confidence intervals. 

\subsubsection{Direct non-parametric bootstrap}
In the direct  bootstrap,  $B$
bootstrap
samples are generated
by resampling GP practices in the observed data with replacement. This is to preserve the correlation structure of observations within the  multi-level clusters. For each bootstrap sample, the parameter estimate
$\hat{\boldsymbol{\theta}}_b$ ($b=1, \ldots, B$) is calculated by solving the score equations ${\bU}_b(\btheta) =
\mathbf{0}$ based on the composite likelihood of the bootstrap sample. The bootstrap estimates of $95\%$ confidence intervals of $\hat{\btheta}$ are given by the  $2.5\%$ and $97.5\%$ sample quantiles of $\hat{\boldsymbol{\theta}}_b$.

\subsubsection{One-step estimating function bootstrap}
The estimating function  bootstrap (EFB) was proposed by  \cite{Hu2000} as an alternative to the direct bootstrap. The main advantages of the EFB
 over the direct bootstrap are computational efficiency and accuracy. The estimating function is  only solved once,  rather than $B+1$ times,  using the original observed data. This  can reduce computing time considerably, especially when iterative procedures are used in the estimation. In our case, the EFB is particularly attractive because of the amount of multi-level data in the CPRD analysis.  
Furthermore, \cite{Binder2004} found that, when using the direct bootstrap for logistic regression, it was possible to have several bootstrap samples for which the parameter estimation algorithm would not converge due to ill-conditioned matrices that were not invertible. To overcome this problem, \cite{Binder2004} and \cite{Rao2004} extended the EFB  to the  survey sampling setting. In addition, \cite{Roberts2009a} developed the EFB for marginal logistic models with longitudinal survey data. As we use multinomial logistic regressions for the CPRD data and   similar convergence problems  are likely when the direct bootstrap is applied, adapting the EFB to the setting of the CPRD data will improve computational efficiency and accuracy. 

Here we  explain the basic idea of the EFB. In order to obtain parameter estimates from the $b$th bootstrap sample,  we need to solve the score equations ${\bU}_b(\btheta) =
\mathbf{0}$.  If we apply Taylor linearization to the left-hand side of the score equations at $\hat{\btheta}$, we have 
\[
\bU_b(\hat{\btheta})+\left\{\frac{\partial \bU_b ({\btheta})}{\partial \btheta} \right\}_{\btheta=\hat{\btheta}}(\btheta-\hat{\btheta}) \approx \mathbf{0}
\]
After rearranging the terms,
\[
\btheta \approx \hat{\btheta}- \left\{\frac{\partial \bU_b ({\btheta})}{\partial \btheta} \right\}_{\btheta=\hat{\btheta}}^{-1} \bU_b(\hat{\btheta}). 
\]
Therefore, a reasonable parameter estimate from the $b$th bootstrap sample is 
\begin{eqnarray}
\hat{\btheta}_b^{EF} &\approx & \hat{\btheta}- \left\{\frac{\partial \bU ({\btheta})}{\partial \btheta} \right\}_{\btheta=\hat{\btheta}}^{-1} \bU_b(\hat{\btheta}) =\hat{\btheta}- \bSigma(\hat{\btheta}) \bU_b(\hat{\btheta})
\end{eqnarray}
 Note that the inverse matrix $\bSigma(\hat{\btheta})$ is based on  the original sample and thus only needs to be evaluated once. 
 $\bU_b(\hat{\btheta})$ is based on the bootstrap sample and evaluated at $\hat{\btheta}$.  The confidence intervals based on $\hat{\boldsymbol{\theta}}_b^{EF}$ can be constructed similarly as in   the direct bootstrap approach. For the multinomial logistic model in~\eqref{model},  $\bSigma(\hat{\btheta})$ can be obtained from the output of the \texttt{multinom} function 
in the \texttt{nnet} package and   $\bU_b(\hat{\btheta})$  can be easily calculated for each  bootstrap sample. For estimating transition probabilities from no improvement in haemoglobin (State 1) in  the CPRD analysis, it  took 59.6 seconds to obtain  parameter estimates for one bootstrap sample in the direct bootstrap with  4 cores in parallel on a Linux cluster (CPU: Intel Xeon E7-8860 v3, 16GB memory per core), while the EFB took only 3.4 seconds.  In the CPRD analysis and simulations reported in this paper, we use high performance clusters with multiple  cores  and large memories to  speed up the computation.

\subsection{Predicted probability of state occupancy over time}

To provide evidence for  costs and benefits of oral iron prescriptions, e.g., in future economic analyses of the CPRD data, we can use the fitted model in~\eqref{model} to predict probabilities of being in different  disease states at fixed time points following oral iron  prescriptions, given time-invariant covariates such as compound classes and doses of oral iron. This is based on the assumption that at time 0 when treatment starts, a patient is always in State 1 (no improvement in haemoglobin).  Confidence intervals can also be easily constructed by using parameter estimates based on bootstrap samples.

\section{Analysis of the CPRD data}\label{analysis}
In this section, we apply the methods described in Section~\ref{methods} to the CPRD data.  Recall that our interest is to estimate transition probabilities of IDA disease states and examine their associations with covariate factors, including the oral iron compound classes and daily dose of oral iron and socio-economic deprivation level measured by IMD2010 scores. We focus on ferrous fumarate and ferrous sulphate in our analysis. The R code for this analysis is included in the Supplementary Materials.
\subsection{Covariates and adjustment variables}

 Based on clinical input, we categorize the daily iron dose into three categories of iron intake: low daily dose of iron (0, 69mg],  medium daily dose of iron [70mg, 150mg]  and high daily dose of iron (150mg, 300mg]. The courses of treatment with daily dose of iron more than 300mg are excluded as these were  rare and were possibly  due to recording errors since there should not be iron doses above 300mg per day based on clinical input. Oral iron compound classes and IMD2010 deprivation scores (1=least deprived, ..., 5=most deprived) are both treated as categorical variables in the model in~\eqref{model}. We also allow interactions between compound classes and daily doses of iron since these are the two main covariates of interest and it is likely that IDA improvement and treatment intolerance depend on different combinations of iron compound classes and iron doses.

In terms of adjustment variables in the model in~\eqref{model}, we choose their functional forms,  i.e.,  $f_j(\cdot)$, $f_t(\cdot)$ and $f_v(\cdot)$  as follows. Since the majority of patients had fewer than 4 courses of treatment during the study period, we categorize the index of the course of treatment  $j$ as 1, 2, 3, $\ge 4$. Since all patients had at least one follow-up assessment after oral iron  prescription,  there are many observations with  transitions from State 1 at $t=0$. In addition,   the distribution of the follow-up time $t$ (in days) is right skewed.  Therefore we create an indicator variable $I(t=0)$ in the model for transitions from State 1  and also include the log transformation $\tilde{t}=\log(t+1)-4$ and its quadratic term ${\tilde{t}}^2$ in all the models for transition probabilities. For the gap time $v$ (in days), we also use the log transformation $\tilde{v}=\log(v+1)-4$ and its quadratic term
$\tilde{v}^2$ since its distribution is also right skewed (see Figure~\ref{figure2}).

\subsection{Estimates of covariate effects}
Since there are different estimates of regression coefficients depending on the initial states of the transitions, it would be useful and clear to summarize the parameter estimates and 95\% confidence intervals using graphical approaches. 
Figure~\ref{classeffect} presents the regression coefficients and  95\% confidence intervals for the effect of oral iron compound classes  by daily iron doses, conditional on other covariates and adjustment variables.  The 95\% confidence intervals are based on the direct bootstrap with 1000 bootstrap samples, which are almost identical to the confidence intervals constructed from the EFB.

Overall, the point estimates of the effects of oral iron compound classes  given daily iron doses are small, with the range $[0.53, 1.29]$ at the scale of  odds ratio \cite[]{Chen2010}. In addition, there are no clear patterns  in the directions of the effect estimates.  This suggests that there are no huge differences between ferrous sulphate and ferrous fumarate in terms of IDA improvement and treatment intolerance, controlling for other factors. 
Since statistical significance at 5\% level can always be achieved given sufficient sample size, no matter how small the true effect size is, we need to cautiously interpret  the results of statistical significance indicated by 95\% confidence intervals  because of the untraditionally large sample size of the CPRD data.  Moreover, these results can only be interpreted as observed associations, since many confounding factors for iron salt prescriptions and IDA  improvement have not been adjusted for.

\begin{figure}[htp]
\centering
\centering\includegraphics[scale=0.65]{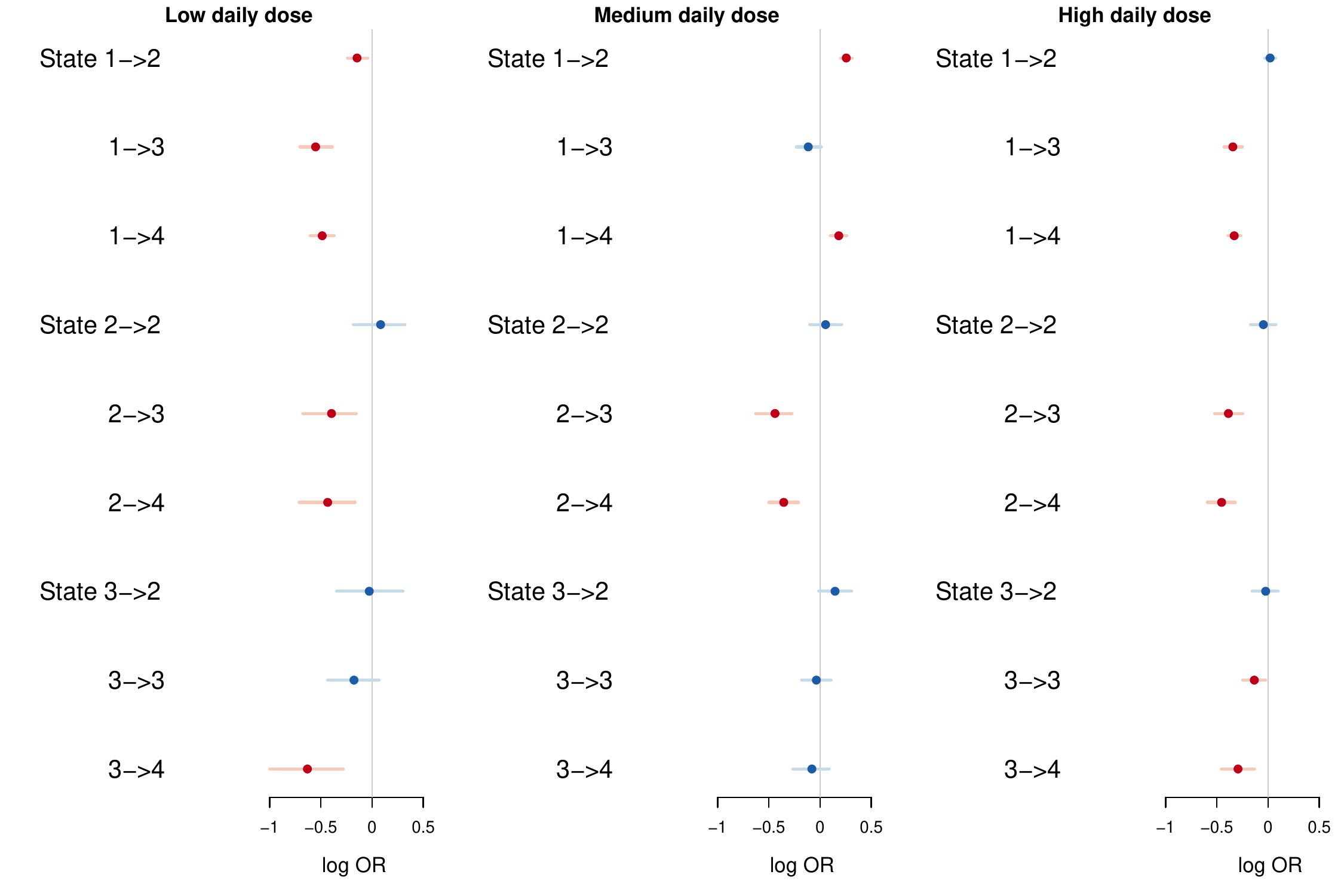}
\caption{Regression coefficient (log odds ratio) estimates and 95\% confidence intervals for the effect of oral iron compound class (ferrous sulphate vs. ferrous fumarate)   by daily iron doses. State 1: no improvement in haemoglobin, 
State 2: improvement in haemoglobin,
State 3:  hospital referral,
State 4: anaemia resolution. Positive (negative) log odd ratios indicate higher (lower) probability of having a specific transition for patients prescribed with ferrous sulphate, compared with patients prescribed with ferrous fumarate. The
estimated log odd ratios with 95\% confidence intervals covering zero (i.e. statistically non-significant)  and not covering zero (i.e. statistically significant) are in blue and red, respectively.}\label{classeffect}
\end{figure}

It would  also be interesting to examine the effects of the adjustment variables, e.g., the gap time. Figure~\ref{gapeffect} plots the estimated curves for the gap time effects by different transitions. It is clear that longer gap times are associated with  higher probabilities of transitioning into improvement in haemoglobin and anaemia resolution.  Shorter gap times are associated with  higher probabilities of hospital referral. This confirms our  conjecture about the informativeness of the visiting process as discussed in Section~\ref{Intro}.

\begin{figure}[htp]
\centering
\centering\includegraphics[scale=0.55]{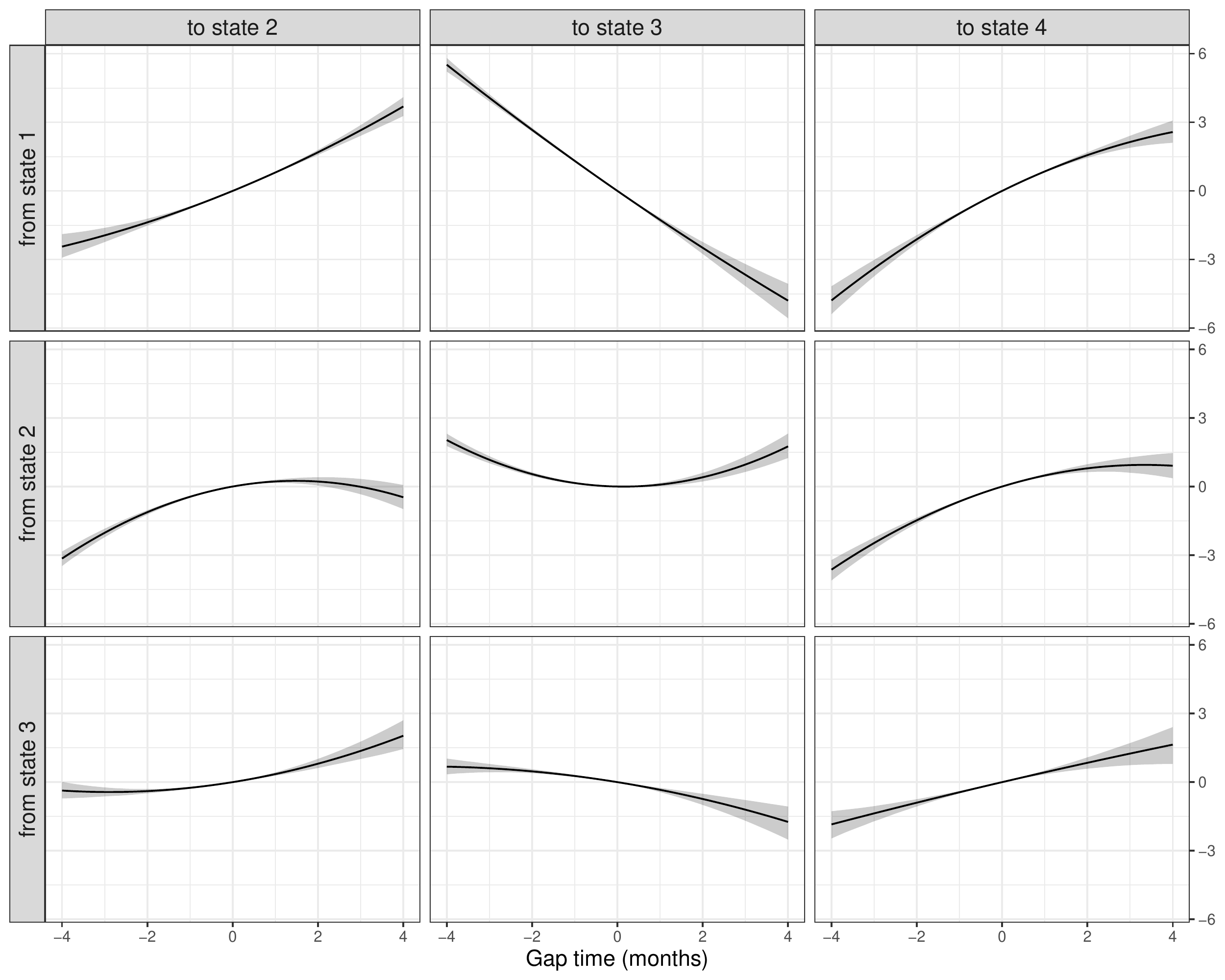}
\caption{Effect of gap time  and 95\% point-wise confidence intervals. State 1: no improvement in haemoglobin, 
State 2: improvement in haemoglobin,
State 3:  hospital referral,
State 4: anaemia resolution. }\label{gapeffect}
\end{figure}

\subsection{State occupancy probabilities}
Figure~\ref{predprob} presents the barplots of the predicted probabilities of  being in different  disease states every two months following oral iron prescriptions, stratified by courses of treatment and iron compound classes, given medium daily dose of oral iron and medium social-economic deprivation level (IMD2010=3). Again it appears that there are no large differences in term of predicted  probabilities of state occupancy between ferrous sulphate  and ferrous fumarate.
This is consistent with findings about small effects of oral iron compound classes in Figure~\ref{classeffect}.
The predicted probabilities of state occupancy can feed into subsequent health economic analyses for oral irons.

\subsection{Other results}
The complete  results of the regression coefficients and both types of  95\% confidence intervals as well as the plots for the effects of daily iron doses by oral iron compound classes and the IMD2010 scores can be found in Section 2 of the Supplementary Materials. Overall, the point estimates of the iron dose effects given iron compound classes are small, with the range  $[0.59, 1.17]$ at the scale of  odds ratio. Again, no clear patterns can be found for the directions of these point estimates.  
The estimated effect sizes for IMD2010 deprivation scores are also small, with the range  $[0.63,1.16]$ for the  odd ratios. However, we can see that  patients from  more deprived regions (IMD2010=4 or 5) were  less likely to  move into improvement in haemoglobin and anaemia resolution than patients from the least deprived regions (IMD2010=1), regardless of their initial states for transitioning. This phenomenon is most prominent for the patients from the most deprived region (i.e., IMD2010=5), and slightly reduced for the patients with  IMD2010=4.

Overall, we found that there is no enough evidence that IDA improvement and treatment intolerance were associated with oral iron compounds ferrous sulphate and ferrous fumarate and their doses, but IDA disease progression may be related to patients' socio-economic status. 

\begin{figure}[htp]
\centering
\centering\includegraphics[scale=0.55]{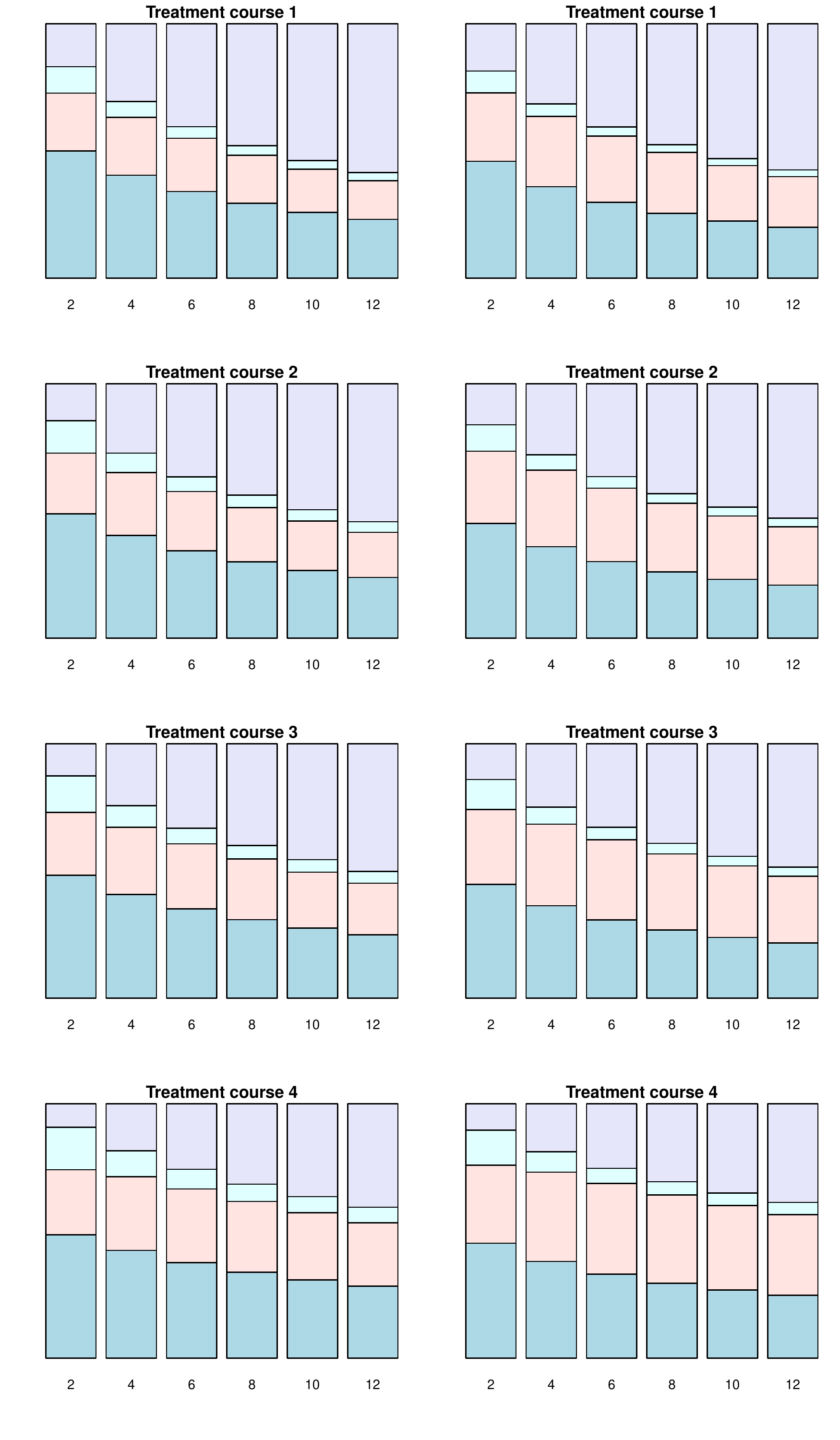}
\caption{Predicted probabilities of state occupancy  every two months following oral iron prescriptions, given medium daily doses of oral iron prescriptions and medium social-economic deprivation level (IMD2010=3). Left panels are barplots for ferrous fumarate;  right panels are barplots for ferrous sulphate. 
\protect\circleblue~ no improvement in haemoglobin; \protect\circlerose ~improvement in haemoglobin; \protect\circlecyan ~hospital referral; \protect\circlelav ~anaemia resolution.
}\label{predprob}
\end{figure}

\section{Simulation Study}\label{simulation}
We conduct a simulation study, with the design motivated by  the CPRD data analysis,  to evaluate the performance of  bootstrap methods in constructing confidence intervals for parameters in the discrete-time Markov model. Details for the simulations can be found in Section 3 of  the Supplementary Materials.
For simplicity, we only consider two disease states (State 1: no improvement of haemoglobin; State 2:  improvement of haemoglobin). Eight hundred datasets are generated based on logistic models of transition probabilities with random effects for GP practices and patients. Overall, in all scenarios with different correlation parameter values for patient-level random effects, the biases of regression coefficient estimators for the discrete-time Markov model are negligible. Reassuringly, confidence intervals from  both direct bootstrap and EFB have good coverages. The correlation parameter has minimal impact on the performance of the point estimator and  the bootstrap methods.

\section{Conclusion and discussion}\label{conclusion}

Motivated by the challenges in analysing the CPRD data for investigating IDA disease progression following oral iron prescriptions, we have developed a computationally efficient approach to modelling disease progression using EHRs data, while accounting for variability at multi-level clusters and informative observation times.  
Our modelling approach is straightforward for practical implementation since no joint model with random effects and no   specialised software are required. Together with the computationally efficient EFB, the proposed approach offers a feasible solution to the   analysis and inference of the motivating CPRD data, which feeds into the subsequent economic analysis of oral irons that is currently in progress.

While our methods are directly motivated by challenges we encountered when analyzing the CPRD data for answering  the IDA research question, they can be readily adapted to other settings with multi-level data of large volume and informative observation times. For example, when fitting random effects model to multi-level data is time-consuming, the EBF can be employed to provide an efficient alternative to direct non-parametric bootstrap. Our conditioning approach for addressing informative observation times can also be incorporated in more complex transition models for longitudinal data \cite[]{Zeng2007}.

The results of the CPRD data analysis should be interpreted in light of several important limitations. First is the potential measurement error in ascertaining the IDA disease states. Specifically, improvement in haemoglobin was not determined exclusively by  haemoglobin test results but also based on medical codes provided by the CPRD. 
Therefore, misclassification error is likely regarding  improvement and no improvement of patients'  haemoglobin levels.
In addition, hospital referrals included any hospital referrals recorded in  HES, such as gynaecology, oncology, gastroenterology, women's clinic, pharmacy, dietician, obstetrics, Accident \& Emergency, heart failure, etc.. As a result, these hospital referrals were not necessarily manifestations of patients' intolerance to iron salt prescriptions. 
This `phenotyping' problem for disease outcomes is a common challenge in analyses using EHRs data.

Second, only a few covariates that are of scientific interests were included in the discrete-time Markov model, and information about other factors that might influence both iron salt prescriptions and  patients' IDA improvement and intolerance (e.g., co-morbidities including gastrointestinal disorders) was not available therefore not examined. As a result, we emphasize that the results in Section~\ref{analysis} do not have causal interpretations and can only be interpreted as observed associations.  In addition, the small set of covariates and adjustment variables included in the CPRD analysis  probably  made the large sample size of the CPRD data be a  
main factor driving the statistical significance shown by 95\% confidence intervals, while the point estimates of covariate effects are small.  Therefore, we have to cautiously interpret the results based on 95\% confidence intervals in relation to point estimates. 
Third, more flexible functional forms for the adjustment variables could be adopted. This will involve model selection procedures to avoid over-fitting. It warrants further research on how to conduct model selection efficiently with large amount of multi-level data.

\begin{center}
{\bf{\large Supplementary Materials}}
\end{center}
The reader is referred to Supplementary Materials for additional results of the CPRD data analysis and details of the simulation study. 

\begin{center}

{\bf{\large Acknowledgements}}
\end{center}
The authors would like to thank Dr Dyfrig Hughes at Bangor University for helpful comments.
The CPRD and HES data extraction was funded through an MRCT Technology DGF grant A853/0174.
The authors were supported by the Medical Research Council [Unit Programme number MC\_UU\_00002/8 and MR/R005699/1].

\vspace{0.2in}
\textit{Conflicts of interest}:
D.I.A.P. has since moved to full employment with Vifor Pharma UK. 
Notwithstanding, the authors declare no potential conflicts of interest with respect to the research, authorship, and/or publication of this article.

\bibliographystyle{rss}

\bibliography{jmref}

\begin{thebibliography}{23}
\expandafter\ifx\csname natexlab\endcsname\relax\def\natexlab#1{#1}\fi
\expandafter\ifx\csname url\endcsname\relax
  \def\url#1{\texttt{#1}}\fi
\expandafter\ifx\csname urlprefix\endcsname\relax\def\urlprefix{URL }\fi

\bibitem[{Binder \emph{et~al.}(2004)Binder, Kovacevic and Roberts}]{Binder2004}
Binder, D.~A., Kovacevic, M. and Roberts, G. (2004) {Design-based methods for
  survey data: alternative uses of estimating functions}.
\newblock In \emph{Proceedings of the Survey Research Methods Section, American
  Statistical Association, American Statistical Association, Washington, DC},
  3301--3312.

\bibitem[{Chen \emph{et~al.}(2010)Chen, Cohen and Chen}]{Chen2010}
Chen, H., Cohen, P. and Chen, S. (2010) How big is a big odds ratio?
  {I}nterpreting the magnitudes of odds ratios in epidemiological studies.
\newblock \emph{Communications in Statistics - Simulation and Computation},
  \textbf{39}, 860--864.

\bibitem[{Cook and Lawless(2007)}]{Cook2007}
Cook, R. and Lawless, J. (2007) \emph{The Statistical Analysis of Recurrent
  Events}.
\newblock New York: Springer-Verlag.

\bibitem[{Daniels and Hogan(2008)}]{Daniels2008}
Daniels, M. and Hogan, J. (2008) \emph{Missing Data in Longitudinal Studies:
  Strategies for Bayesian Modeling and Sensitivity Analysis.}, vol. 101 of
  \emph{Monographs on Statistics and Applied Probability}.
\newblock New York: {C}hapman \& {H}all/{CRC}.

\bibitem[{Herrett \emph{et~al.}(2015)Herrett, Gallagher, Bhaskaran, Forbes,
  Mathur, van Staa and Smeeth}]{Herrett2015}
Herrett, E., Gallagher, A.~M., Bhaskaran, K., Forbes, H., Mathur, R., van Staa,
  T. and Smeeth, L. (2015) {Data Resource Profile: Clinical Practice Research
  Datalink (CPRD)}.
\newblock \emph{International Journal of Epidemiology}, \textbf{44}, 827--836.
\newblock \urlprefix\url{https://doi.org/10.1093/ije/dyv098}.

\bibitem[{Hu and Kalbfleisch(2000)}]{Hu2000}
Hu, F. and Kalbfleisch, J. (2000) {The estimating function bootstrap}.
\newblock \emph{The Canadian Journal of Statistics}, \textbf{28}, 449--481.

\bibitem[{Lange \emph{et~al.}(2015)Lange, Hubbard, Inoue and Minin}]{Lange2014}
Lange, J.~M., Hubbard, R.~A., Inoue, L. Y.~T. and Minin, V.~N. (2015) {A joint
  model for multistate disease processes and random informative observation
  times, with applications to electronic medical records data}.
\newblock \emph{Biometrics}, \textbf{71}, 90--101.

\bibitem[{Lin and Ying(2001)}]{Lin2001}
Lin, D. and Ying, Z. (2001) Semiparametric and nonparametric regression
  analysis of longitudinal data.
\newblock \emph{J Am Stat Assoc}, \textbf{96}, 103--113.

\bibitem[{Lin \emph{et~al.}(2004)Lin, Scharfstein and Rosenheck}]{Lin2004}
Lin, H., Scharfstein, D. and Rosenheck, R. (2004) Analysis of longitudinal data
  with irregular, outcome-dependent follow-up.
\newblock \emph{J R Stat Soc Ser B}, \textbf{66}, 791--813.

\bibitem[{Lindgren \emph{et~al.}(2009)Lindgren, Wikman, Befrits, Blom,
  Eriksson, Gr\"{a}nn\"{o}, Ung, Hjortswang, Lindgren and Unge}]{Lindgren2009}
Lindgren, S., Wikman, O., Befrits, R., Blom, H., Eriksson, A., Gr\"{a}nn\"{o},
  C., Ung, K., Hjortswang, H., Lindgren, A. and Unge, P. (2009) Intravenous
  iron sucrose is superior to oral iron sulphate for correcting anaemia and
  restoring iron stores in ibd patients: A randomized, controlled,
  evaluator-blind, multicentre study.
\newblock \emph{Scandinavian journal of gastroenterology}, \textbf{44(7)},
  838--845.

\bibitem[{{Public Health England}(2018)}]{NDNS}
{Public Health England} (2018) National diet and nutrition survey: Results from
  years 7 and 8 (combined) of the rolling programme (2014/2015--2015/2016).
\newblock
  \urlprefix\url{https://www.gov.uk/government/statistics/ndns-results-from-years-7-and-8-combined}.

\bibitem[{Pullenayegum and Lim(2016)}]{Pull2016}
Pullenayegum, E. and Lim, L. (2016) Longitudinal data subject to irregular
  observation: A review of methods with a focus on visit processes,
  assumptions, and study design.
\newblock \emph{Statistical Methods in Medical Research}, \textbf{25},
  2992--3014.

\bibitem[{Rabe-Hesketh and Skrondal(2006)}]{Rabe-Hesketh2006}
Rabe-Hesketh, S. and Skrondal, A. (2006) Multi-level modelling of complex
  survey data.
\newblock \emph{Journal of the Royal Statistical Society, Series C},
  \textbf{169}, 805--827.

\bibitem[{Rao and Tausi(2004)}]{Rao2004}
Rao, J. and Tausi, M. (2004) {Estimating function jackknife variance estimators
  under stratified multistage sampling}.
\newblock \emph{Communications in Statistics - Theory and Methods},
  \textbf{33}, 2087--2095.

\bibitem[{Roberts \emph{et~al.}(2009)Roberts, Ren and Rao}]{Roberts2009a}
Roberts, G., Ren, Q. and Rao, J. (2009) \emph{Methodology of Longitudinal
  Surveys}, chap. {Using marginal mean models for data from longitudinal
  surveys with a complex design: some advances in methods},  351--366.
\newblock John Wiley \& Sons.

\bibitem[{Saha \emph{et~al.}(2007)Saha, Pandhi, Gopalan, Malhotra and
  Saha}]{Saha2007}
Saha, L., Pandhi, P., Gopalan, S., Malhotra, S. and Saha, P. (2007) Comparison
  of efficacy, tolerability, and cost of iron polymaltose complex with ferrous
  sulphate in the treatment of iron deficiency anemia in pregnant women.
\newblock \emph{Medscape general practice}, \textbf{9(1)}, 1.

\bibitem[{Souza \emph{et~al.}(2009)Souza, Batista, Bresani, Ferreira and
  Figueiroa}]{Souza2009}
Souza, A., Batista, F.~M., Bresani, C., Ferreira, L. and Figueiroa, J. (2009)
  Adherence and side effects of three ferrous sulfate treatment regimens on
  anemic pregnant women in clinical trials.
\newblock \emph{Cadernos de Sa\'{u}de P\'{u}blica}, \textbf{25(6)}, 1225--1233.

\bibitem[{Sun \emph{et~al.}(2007)Sun, Sun and Liu}]{Sun2007}
Sun, J., Sun, L. and Liu, D. (2007) Regression analysis of longitudinal data in
  the presence of informative observation and censoring times.
\newblock \emph{J Am Stat Assoc}, \textbf{102}, 1397--1406.

\bibitem[{Tolkien \emph{et~al.}(2015)Tolkien, Stecher, Mander, Pereira and
  Powell}]{Tolkien2015}
Tolkien, Z., Stecher, L., Mander, A.~P., Pereira, D. I.~A. and Powell, J.~J.
  (2015) Ferrous sulfate supplementation causes significant gastrointestinal
  side-effects in adults: A systematic review and meta-analysis.
\newblock \emph{PLOS ONE}, \textbf{10}, 1--20.
\newblock \urlprefix\url{https://doi.org/10.1371/journal.pone.0117383}.

\bibitem[{Varin \emph{et~al.}(2011)Varin, Reid and Firth}]{Varin2011}
Varin, C., Reid, N. and Firth, D. (2011) {An overview of composite likelihood
  methods}.
\newblock \emph{Statistica Sinica}, \textbf{21}, 5--42.

\bibitem[{Venables and Ripley(2002)}]{Venables2002}
Venables, W.~N. and Ripley, B.~D. (2002) \emph{Modern Applied Statistics with
  S}.
\newblock New York: Springer, fourth edn.
\newblock \urlprefix\url{http://www.stats.ox.ac.uk/pub/MASS4}.
\newblock ISBN 0-387-95457-0.

\bibitem[{Zaim \emph{et~al.}(2011)Zaim, Piselli, Fioravant and
  Kanony-Truc}]{Zaim2011}
Zaim, M., Piselli, L., Fioravant, P. and Kanony-Truc, C. (2011) Efficacy and
  tolerability of a prolonged release ferrous sulphate formulation in iron
  deficiency anaemia: a non-inferiority controlled trial.
\newblock \emph{European Journal of Nutrition}, \textbf{51(2)}, 221--229.

\bibitem[{Zeng and Cook(2007)}]{Zeng2007}
Zeng, L. and Cook, R.~J. (2007) Transition models for multivariate longitudinal
  binary data.
\newblock \emph{Journal of the American Statistical Association}, \textbf{102},
  211--223.

\end{thebibliography}

\end{document}


\maketitle

\baselineskip=24pt

\section{Further details of the CPRD data}

\begin{figure}[htp]
\centering
\centering\includegraphics[scale=0.55]{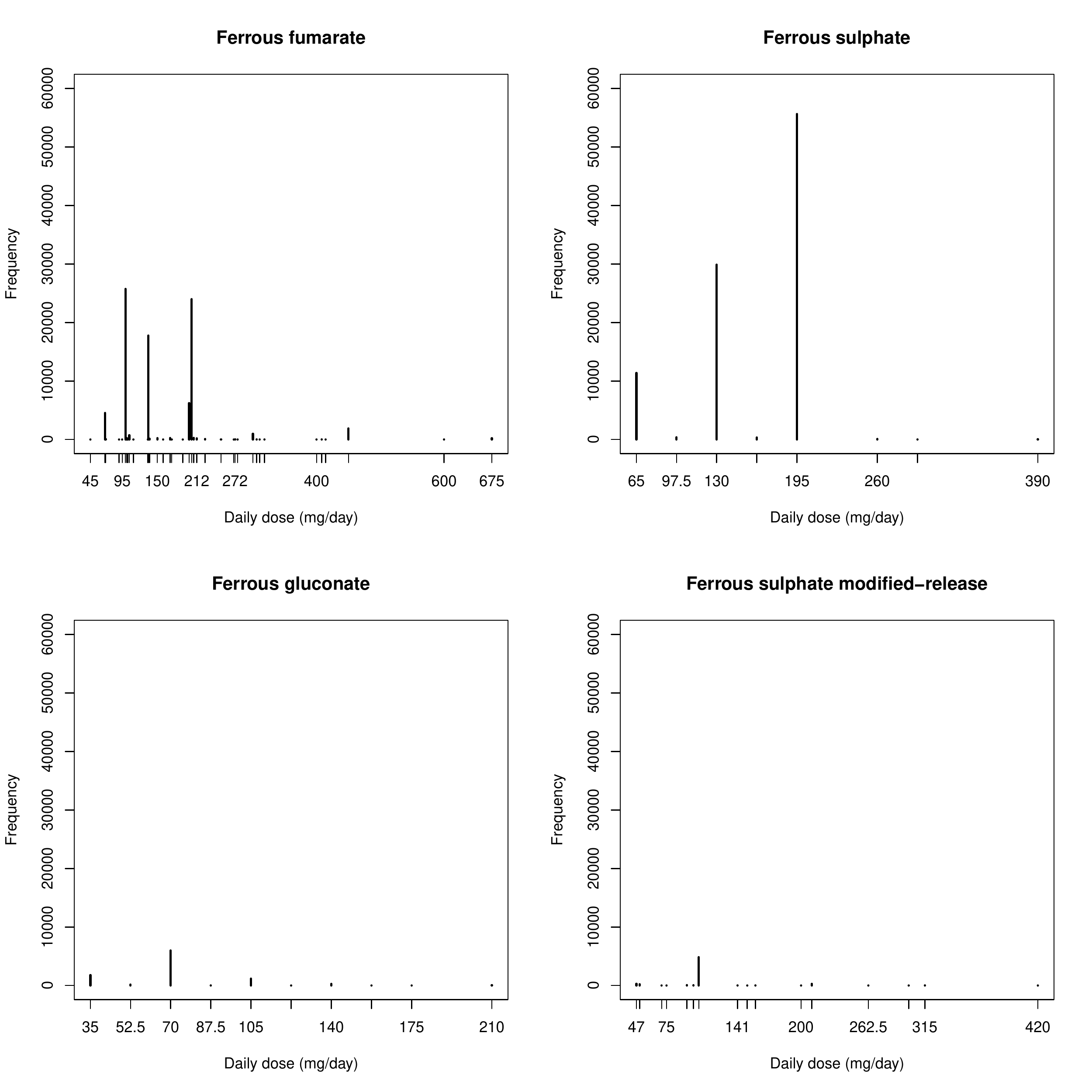}
\caption{Frequency  of average daily doses of iron by iron salt classes prescribed in all courses of treatment  included in the CPRD data.}\label{figure1}
\end{figure}

Figure~\ref{figure1} shows the frequencies of average daily doses of iron by iron salt compound classes for all courses of treatment.  
The daily iron doses are clustered at several levels due to  specific instructions in the prescriptions about  the number of pills taken per day and the iron dosing in each pill.  %

From Table 2 of the main text, it can be seen that there were about 1/3 of the patients whose IMD2010 data were not available at the time of CPRD data extraction and linkage. This is probably because it was not possible to link the postcodes of these patients' registered addresses at GP practices  to  calculated Index of Deprivation  at the level of lower layer super output area (LSOA, a geospatial statistical unit used in England and Wales to facilitate the reporting of small area statistics). In our analysis, we include a separate category for the missing IMD2010.

\section{Additional results of the CPRD data analysis}\label{analysis}

\subsection{Effects of daily iron dose by iron classes}
\begin{figure}[htp]
\centering
\centering\includegraphics[scale=0.65]{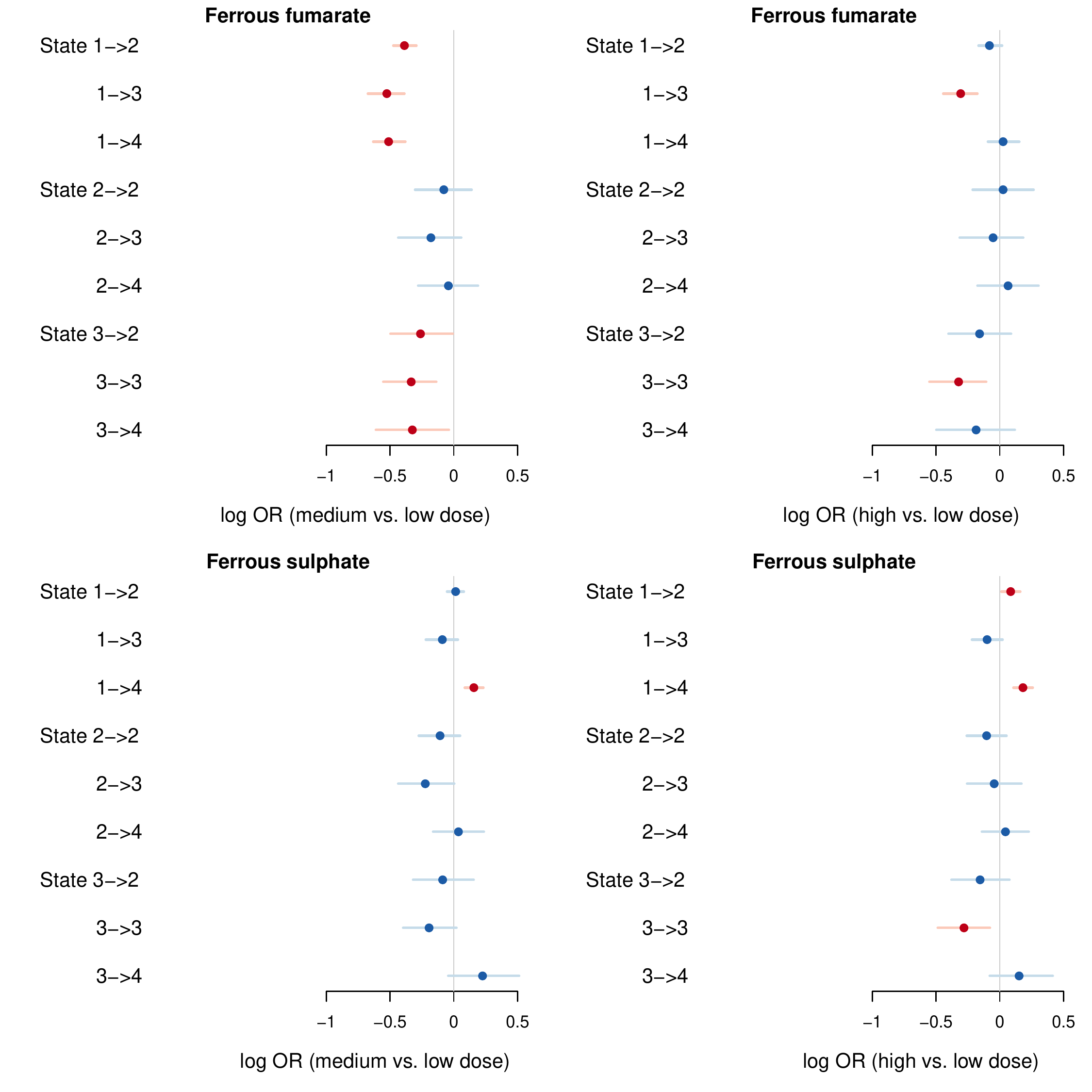}
\caption{Daily iron dose effect estimates and 95\% confidence intervals, conditional on the iron compound class. State 1: no improvement in haemoglobin, 
State 2: improvement in haemoglobin,
State 3:  hospital referral,
State 4: anaemia resolution. The
estimated log odd ratios with 95\% confidence intervals covering zero (i.e. statistically non-significant)  and not covering zero (i.e. statistically significant) are highlighted in blue and red, respectively. }\label{doseeffect}
\end{figure}

Figure~\ref{doseeffect} presents the regression coefficients and 95\% confidence intervals (based on direct bootstrap)  for the daily iron dose effects by iron compound classes. 
Overall, the point estimates of the iron dose effects are small, with the range $[0.59, 1.17]$ at the scale of  odds ratio. Again, no clear patterns can be found for the directions of these point estimates.

\subsection{Effects of IMD2010 deprivation scores}

\begin{figure}[htp]
\centering
\centering\includegraphics[scale=0.55]{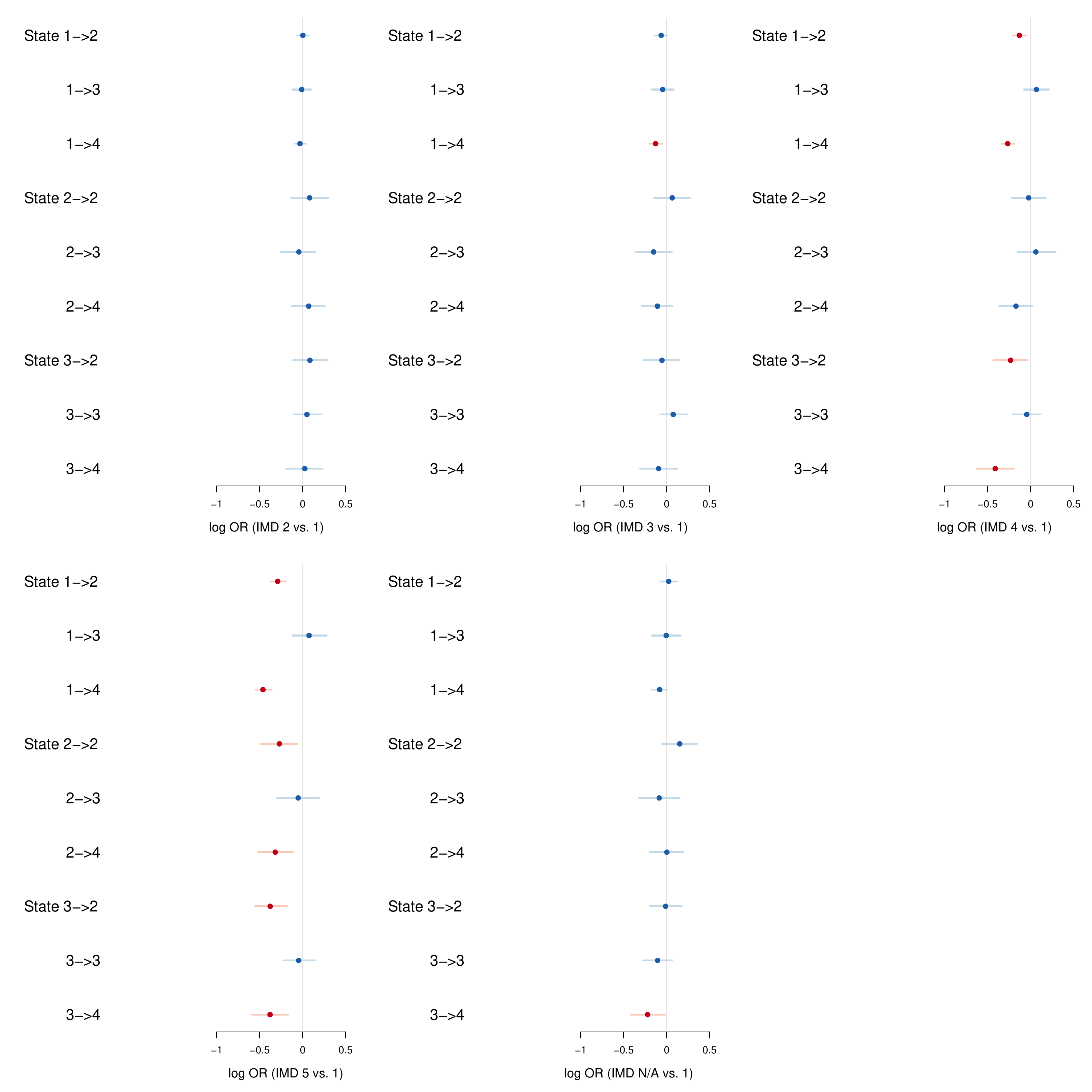}
\caption{Effect estimates of IMD2010 scores and 95\%  confidence intervals. State 1: no improvement in haemoglobin, 
State 2: improvement in haemoglobin,
State 3:  hospital referral,
State 4: anaemia resolution. The
estimated log odd ratios with 95\% confidence intervals covering zero (i.e. statistically non-significant)  and not covering zero (i.e. statistically significant) are highlighted in blue and red, respectively.}\label{imdeffect}
\end{figure}

Figure~\ref{imdeffect} presents the regression coefficients and 95\% confidence intervals (based on direct bootstrap) for the effects
of socio-economic deprivation level measured by IMD2010 scores. The estimated effect sizes for IMD2010 deprivation scores are small, with the range  $[0.63,1.16]$ for the  odd ratios. However, we can see that  patients from  more deprived regions (IMD2010=4 or 5) were  less likely to  move into improvement in haemoglobin and anaemia resolution than patients from the least deprived regions (IMD2010=1), regardless of their initial states for transitioning. This phenomenon is most prominent for the patients from the most deprived region (i.e. IMD2010=5), and slightly reduced for the patients with  IMD2010=4.

Tables~\ref{tablestate0}-\ref{tablestate4} present all regression coefficient estimates and 95\% confidence intervals based on both direct bootstrap and estimating function bootstrap (EFB). The confidence intervals generated by the two bootstrap methods are almost identical.

\begin{sidewaystable}[!ht]
\caption{Regression coefficient estimates and 95\% confidence intervals by estimating function bootstrap (EFB) and direct bootstrap (DB)  for transition probabilities from State 1 (no improvement of haemoglobin) to other states in the CPRD data. Iron class 3: ferrous sulphate; iron class 1: ferrous fumerate.}
\label{tablestate0}
\centering
\footnotesize{\begin{tabular}{lrrrrrrrrrrrrrrr}
  \hline
 & \multicolumn{5}{c}{\textbf{State 1 $\rightarrow$ State 2} } & \multicolumn{5}{c}{\textbf{State 1 $\rightarrow$ State 3} }& \multicolumn{5}{c}{\textbf{State 1 $\rightarrow$ State 4} }\\ 
& \multicolumn{1}{c}{\scriptsize{Point}} &\multicolumn{2}{c}{EFB }& \multicolumn{2}{c}{DB } & \multicolumn{1}{c}{\scriptsize{Point}} &\multicolumn{2}{c}{EFB }& \multicolumn{2}{c}{DB } & \multicolumn{1}{c}{\scriptsize{Point}} &\multicolumn{2}{c}{EFB }& \multicolumn{2}{c}{DB }\\ 
 & \scriptsize{estimate}  & \scriptsize{2.5\%} & \scriptsize{97.5}\% & \scriptsize{2.5\%} & \scriptsize{97.5}\% & \scriptsize{estimate}  & \scriptsize{2.5\%} & \scriptsize{97.5}\%& \scriptsize{2.5\%} & \scriptsize{97.5}\% & \scriptsize{estimate} & \scriptsize{2.5\%} & \scriptsize{97.5}\% & \scriptsize{2.5\%} & \scriptsize{97.5}\% \\ 
  \hline
Intercept    & -0.97 & -1.07 & -0.86 & -1.08 & -0.86 & -3.74 & -3.95 & -3.53 & -3.95 & -3.53 & -1.25 & -1.37 & -1.12 & -1.38 & -1.12 \\ 
Treatment: $\mbox{I}(j=2)$ & 0.07 & 0.04 & 0.10 & 0.04 & 0.10 & 0.22 & 0.18 & 0.25 & 0.18 & 0.25 & -0.12 & -0.16 & -0.09 & -0.16 & -0.09 \\ 
~~~~~~~~~~~~~~~ $\mbox{I}(j=3)$ & 0.12 & 0.08 & 0.16 & 0.08 & 0.16 & 0.35 & 0.30 & 0.40 & 0.30 & 0.40 & -0.25 & -0.30 & -0.21 & -0.30 & -0.21 \\ 
~~~~~~~~~~~~~~~ $\mbox{I}(j\ge 4)$ & 0.15 & 0.10 & 0.19 & 0.10 & 0.19 & 0.50 & 0.44 & 0.56 & 0.44 & 0.56 & -0.57 & -0.62 & -0.52 & -0.62 & -0.52 \\ 
         Time:~~~~~~~ $\mbox{I}(t_{gijk}=0)$ & -0.01 & -1.02 & 0.99 & -1.01 & 1.05 & -0.67 & -1.93 & 0.67 & -2.14 & 0.54 & 2.30 & 0.87 & 3.95 & 0.78 & 3.87 \\ 
 ~~~~~~~~~~~~~~~~$\tilde{t}_{gijk}$ & -0.17 & -0.26 & -0.09 & -0.25 & -0.08 & -0.16 & -0.28 & -0.03 & -0.29 & -0.04 & -0.33 & -0.43 & -0.21 & -0.44 & -0.22 \\ 
  ~~~~~~~~~~~~~~~~$(\tilde{t}_{gijk})^2$ & -0.01 & -0.05 & 0.04 & -0.05 & 0.03 & 0.18 & 0.13 & 0.24 & 0.13 & 0.24 & -0.18 & -0.26 & -0.11 & -0.26 & -0.11 \\ 
 Gap time:~~~$\tilde{v}_{gijk}$ & 0.77 & 0.73 & 0.80 & 0.73 & 0.80 & -1.29 & -1.37 & -1.22 & -1.36 & -1.22 & 0.92 & 0.88 & 0.96 & 0.88 & 0.96 \\ 
  ~~~~~~~~~~~~~~~~$(\tilde{v}_{gijk})^2$ & 0.04 & 0.01 & 0.07 & 0.01 & 0.07 & 0.02 & -0.01 & 0.05 & -0.01 & 0.05 & -0.07 & -0.10 & -0.04 & -0.10 & -0.04 \\ 
 Iron class:~~ 3 vs. 1 & -0.15 & -0.25 & -0.05 & -0.24 & -0.04 & -0.55 & -0.71 & -0.40 & -0.71 & -0.39 & -0.49 & -0.61 & -0.37 & -0.61 & -0.37 \\ 
 Daily iron dose: medium  & -0.39 & -0.48 & -0.30 & -0.47 & -0.29 & -0.53 & -0.66 & -0.38 & -0.67 & -0.39 & -0.51 & -0.65 & -0.40 & -0.63 & -0.38 \\ 
 ~~~~~~~~~~~~~~~~~~~~~ high  & -0.08 & -0.18 & 0.00 & -0.17 & 0.02 & -0.31 & -0.45 & -0.17 & -0.44 & -0.18 & 0.03 & -0.10 & 0.14 & -0.09 & 0.15 \\ 
  Class 3 * medium iron  dose   & 0.40 & 0.30 & 0.52 & 0.29 & 0.50 & 0.44 & 0.25 & 0.62 & 0.25 & 0.62 & 0.67 & 0.53 & 0.82 & 0.52 & 0.81 \\ 
  Class 3 * high iron dose   & 0.17 & 0.07 & 0.27 & 0.06 & 0.26 & 0.21 & 0.05 & 0.38 & 0.03 & 0.37 & 0.16 & 0.03 & 0.28 & 0.04 & 0.29 \\ 
  IMD2010:~~ 1 &  &  &  &  &  &  &  &  &  &  &  &  &  &  &  \\ 
  ~~~~~~~~~~~~~~~~2 & 0.00 & -0.07 & 0.07 & -0.07 & 0.07 & -0.01 & -0.12 & 0.10 & -0.12 & 0.10 & -0.03 & -0.10 & 0.03 & -0.09 & 0.04 \\ 
         ~~~~~~~~~~~~~~~~3 & -0.06 & -0.14 & 0.01 & -0.14 & 0.01 & -0.05 & -0.17 & 0.09 & -0.17 & 0.08 & -0.13 & -0.21 & -0.06 & -0.20 & -0.05 \\ 
         ~~~~~~~~~~~~~~~~4 & -0.13 & -0.21 & -0.06 & -0.21 & -0.06 & 0.07 & -0.09 & 0.22 & -0.08 & 0.21 & -0.27 & -0.34 & -0.20 & -0.34 & -0.19 \\ 
         ~~~~~~~~~~~~~~~~5 & -0.29 & -0.38 & -0.21 & -0.37 & -0.20 & 0.07 & -0.13 & 0.27 & -0.12 & 0.28 & -0.46 & -0.56 & -0.37 & -0.55 & -0.36 \\ 
 ~~~~~~~~~~~~~~~~NA & 0.02 & -0.07 & 0.12 & -0.07 & 0.12 & -0.01 & -0.17 & 0.16 & -0.17 & 0.16 & -0.08 & -0.17 & 0.01 & -0.17 & 0.01 \\ 
   \hline
\end{tabular}}
\end{sidewaystable}

\begin{sidewaystable}[!ht]
\caption{Regression coefficient estimates and 95\% confidence intervals by estimating function bootstrap (EFB) and direct bootstrap (DB)  for transition probabilities from State 2 (improvement of haemoglobin) to other states in the CPRD data. Iron class 3: ferrous sulphate; iron class 1: ferrous fumerate.  }
\label{tablestate1}
\centering
\footnotesize{\begin{tabular}{lrrrrrrrrrrrrrrr}
  \hline
 & \multicolumn{5}{c}{\textbf{State 2 $\rightarrow$ State 2} } & \multicolumn{5}{c}{\textbf{State 2 $\rightarrow$ State 3} }& \multicolumn{5}{c}{\textbf{State 2 $\rightarrow$ State 4} }\\ 
& \multicolumn{1}{c}{\scriptsize{Point}} &\multicolumn{2}{c}{EFB }& \multicolumn{2}{c}{DB } & \multicolumn{1}{c}{\scriptsize{Point}} &\multicolumn{2}{c}{EFB }& \multicolumn{2}{c}{DB } & \multicolumn{1}{c}{\scriptsize{Point}} &\multicolumn{2}{c}{EFB }& \multicolumn{2}{c}{DB }\\ 
 & \scriptsize{estimate}  & \scriptsize{2.5\%} & \scriptsize{97.5}\% & \scriptsize{2.5\%} & \scriptsize{97.5}\% & \scriptsize{estimate}  & \scriptsize{2.5\%} & \scriptsize{97.5}\%& \scriptsize{2.5\%} & \scriptsize{97.5}\% & \scriptsize{estimate} & \scriptsize{2.5\%} & \scriptsize{97.5}\% & \scriptsize{2.5\%} & \scriptsize{97.5}\% \\ 
  \hline
Intercept & 1.17 & 0.89 & 1.44 & 0.90 & 1.46 & -0.30 & -0.60 & -0.01 & -0.59 & -0.01 & 0.74 & 0.48 & 1.00 & 0.48 & 1.03 \\ 
 Treatment: $\mbox{I}(j=2)$ & 0.05 & -0.04 & 0.15 & -0.04 & 0.14 & 0.15 & 0.04 & 0.27 & 0.03 & 0.26 & -0.10 & -0.20 & 0.01 & -0.20 & 0.00 \\ 
 ~~~~~~~~~~~~~~~ $\mbox{I}(j=3)$ & 0.13 & 0.02 & 0.24 & 0.02 & 0.25 & 0.16 & 0.02 & 0.30 & 0.01 & 0.29 & -0.18 & -0.31 & -0.04 & -0.31 & -0.05 \\ 
 ~~~~~~~~~~~~~~~ $\mbox{I}(j\ge 4)$ & 0.15 & 0.03 & 0.26 & 0.03 & 0.26 & 0.25 & 0.13 & 0.36 & 0.14 & 0.37 & -0.47 & -0.59 & -0.36 & -0.59 & -0.36 \\ 
  Time:~~~~~~~ $\tilde{t}_{gijk}$ & 0.34 & 0.25 & 0.43 & 0.25 & 0.43 & 0.17 & 0.06 & 0.29 & 0.05 & 0.28 & -0.00 & -0.11 & 0.11 & -0.12 & 0.10 \\ 
  ~~~~~~~~~~~~~~~~$(\tilde{t}_{gijk})^2$ & -0.04 & -0.08 & 0.00 & -0.08 & 0.00 & 0.01 & -0.05 & 0.05 & -0.04 & 0.06 & -0.06 & -0.11 & -0.01 & -0.11 & -0.01 \\ 
  Gap time:~~~$\tilde{v}_{gijk}$ & 0.33 & 0.28 & 0.38 & 0.29 & 0.39 & -0.03 & -0.09 & 0.02 & -0.09 & 0.02 & 0.57 & 0.52 & 0.62 & 0.52 & 0.62 \\ 
  ~~~~~~~~~~~~~~~~$(\tilde{v}_{gijk})^2$ & -0.11 & -0.14 & -0.09 & -0.14 & -0.09 & 0.12 & 0.09 & 0.14 & 0.10 & 0.14 & -0.09 & -0.11 & -0.06 & -0.12 & -0.06 \\ 
  Iron class:~~ 3 vs. 1  & 0.08 & -0.15 & 0.34 & -0.18 & 0.32 & -0.40 & -0.65 & -0.13 & -0.68 & -0.15 & -0.43 & -0.70 & -0.15 & -0.71 & -0.16 \\ 
  Daily iron dose: medium & -0.08 & -0.30 & 0.16 & -0.30 & 0.14 & -0.18 & -0.42 & 0.07 & -0.44 & 0.06 & -0.04 & -0.28 & 0.18 & -0.28 & 0.19 \\ 
 ~~~~~~~~~~~~~~~~~~~~~ high & 0.03 & -0.22 & 0.26 & -0.21 & 0.27 & -0.05 & -0.29 & 0.21 & -0.32 & 0.18 & 0.06 & -0.17 & 0.30 & -0.18 & 0.30 \\ 
  Class 3 * medium iron dose  & -0.03 & -0.29 & 0.23 & -0.30 & 0.23 & -0.04 & -0.35 & 0.25 & -0.33 & 0.26 & 0.08 & -0.24 & 0.39 & -0.23 & 0.39 \\ 
  Class 3 * high iron dose & -0.13 & -0.41 & 0.13 & -0.40 & 0.15 & 0.01 & -0.29 & 0.30 & -0.28 & 0.32 & -0.02 & -0.32 & 0.28 & -0.32 & 0.28 \\ 
 IMD2010:~~~1 &  &  &  &  &  &  &  &  &  &  &  &  &  &  &  \\ 
    ~~~~~~~~~~~~~~~~2 & 0.08 & -0.14 & 0.29 & -0.13 & 0.30 & -0.04 & -0.24 & 0.17 & -0.25 & 0.15 & 0.07 & -0.12 & 0.26 & -0.13 & 0.26 \\ 
  ~~~~~~~~~~~~~~~~3 & 0.06 & -0.15 & 0.27 & -0.15 & 0.27 & -0.15 & -0.36 & 0.06 & -0.36 & 0.06 & -0.11 & -0.28 & 0.07 & -0.29 & 0.06 \\ 
  ~~~~~~~~~~~~~~~~4 & -0.02 & -0.22 & 0.17 & -0.22 & 0.17 & 0.06 & -0.16 & 0.28 & -0.15 & 0.29 & -0.17 & -0.35 & 0.02 & -0.36 & 0.02 \\ 
   ~~~~~~~~~~~~~~~~5 & -0.27 & -0.48 & -0.05 & -0.49 & -0.06 & -0.05 & -0.29 & 0.19 & -0.30 & 0.19 & -0.32 & -0.52 & -0.12 & -0.52 & -0.11 \\ 
  ~~~~~~~~~~~~~~~~NA & 0.15 & -0.06 & 0.35 & -0.05 & 0.35 & -0.09 & -0.32 & 0.15 & -0.33 & 0.15 & 0.00 & -0.19 & 0.20 & -0.20 & 0.19 \\ 
   \hline
\end{tabular}}
\end{sidewaystable}

\begin{sidewaystable}[!ht]
\caption{Regression coefficient estimates and 95\% confidence intervals by estimating function bootstrap (EFB) and direct bootstrap (DB)  for transition probabilities from State 3 (hospital referral) to other states in the CPRD data. Iron class 3: ferrous sulphate; iron class 1: ferrous fumerate.}
\label{tablestate4}
\centering
\footnotesize{\begin{tabular}{lrrrrrrrrrrrrrrr}
  \hline
 & \multicolumn{5}{c}{\textbf{State 3 $\rightarrow$ State 2} } & \multicolumn{5}{c}{\textbf{State 3 $\rightarrow$ State 3} }& \multicolumn{5}{c}{\textbf{State 3 $\rightarrow$ State 4} }\\ 
& \multicolumn{1}{c}{\scriptsize{Point}} &\multicolumn{2}{c}{EFB }& \multicolumn{2}{c}{DB } & \multicolumn{1}{c}{\scriptsize{Point}} &\multicolumn{2}{c}{EFB }& \multicolumn{2}{c}{DB } & \multicolumn{1}{c}{\scriptsize{Point}} &\multicolumn{2}{c}{EFB }& \multicolumn{2}{c}{DB }\\ 
 & \scriptsize{estimate}  & \scriptsize{2.5\%} & \scriptsize{97.5}\% & \scriptsize{2.5\%} & \scriptsize{97.5}\% & \scriptsize{estimate}  & \scriptsize{2.5\%} & \scriptsize{97.5}\%& \scriptsize{2.5\%} & \scriptsize{97.5}\% & \scriptsize{estimate} & \scriptsize{2.5\%} & \scriptsize{97.5}\% & \scriptsize{2.5\%} & \scriptsize{97.5}\% \\ 
  \hline
Intercept & 0.33 & 0.04 & 0.63 & 0.04 & 0.62 & 0.87 & 0.63 & 1.11 & 0.63 & 1.11 & 0.12 & -0.21 & 0.41 & -0.17 & 0.46 \\ 
 Treatment: $\mbox{I}(j=2)$ & -0.02 & -0.13 & 0.09 & -0.14 & 0.09 & -0.00 & -0.10 & 0.10 & -0.10 & 0.10 & -0.15 & -0.28 & -0.02 & -0.28 & -0.03 \\ 
 ~~~~~~~~~~~~~~~ $\mbox{I}(j=3)$ & -0.19 & -0.32 & -0.04 & -0.33 & -0.06 & -0.10 & -0.21 & 0.02 & -0.22 & 0.01 & -0.36 & -0.52 & -0.20 & -0.53 & -0.21 \\ 
  ~~~~~~~~~~~~~~~ $\mbox{I}(j\ge 4)$ & -0.14 & -0.25 & -0.03 & -0.25 & -0.02 & -0.08 & -0.19 & 0.02 & -0.18 & 0.02 & -0.69 & -0.84 & -0.55 & -0.84 & -0.54 \\ 
  Time:~~~~~~~ $\tilde{t}_{gijk}$ & 0.17 & 0.13 & 0.21 & 0.13 & 0.21 & -0.01 & -0.05 & 0.02 & -0.05 & 0.02 & 0.02 & -0.02 & 0.07 & -0.02 & 0.06 \\ 
  ~~~~~~~~~~~~~~~~$(\tilde{t}_{gijk})^2$ & -0.01 & -0.03 & 0.02 & -0.03 & 0.01 & 0.08 & 0.06 & 0.10 & 0.06 & 0.10 & -0.09 & -0.12 & -0.06 & -0.12 & -0.05 \\ 
  Gap time:~~~$\tilde{v}_{gijk}$ & 0.30 & 0.25 & 0.35 & 0.25 & 0.36 & -0.30 & -0.37 & -0.23 & -0.37 & -0.24 & 0.44 & 0.38 & 0.50 & 0.37 & 0.49 \\ 
  ~~~~~~~~~~~~~~~~$(\tilde{v}_{gijk})^2$ & 0.05 & 0.02 & 0.08 & 0.03 & 0.08 & -0.03 & -0.06 & -0.00 & -0.07 & -0.00 & -0.01 & -0.05 & 0.03 & -0.05 & 0.03 \\ 
  Iron class:~~ 3 vs. 1  & -0.03 & -0.36 & 0.29 & -0.35 & 0.30 & -0.18 & -0.43 & 0.07 & -0.43 & 0.07 & -0.63 & -0.99 & -0.26 & -1.00 & -0.28 \\ 
  Daily iron dose: medium & -0.26 & -0.52 & -0.03 & -0.50 & -0.01 & -0.33 & -0.54 & -0.12 & -0.55 & -0.14 & -0.33 & -0.61 & -0.03 & -0.61 & -0.04 \\ 
 ~~~~~~~~~~~~~~~~~~~~~ high & -0.16 & -0.41 & 0.08 & -0.40 & 0.09 & -0.32 & -0.54 & -0.10 & -0.55 & -0.11 & -0.19 & -0.50 & 0.12 & -0.50 & 0.12 \\ 
  Class 3 * medium iron dose  & 0.17 & -0.19 & 0.57 & -0.20 & 0.53 & 0.14 & -0.12 & 0.43 & -0.14 & 0.41 & 0.55 & 0.14 & 0.95 & 0.15 & 0.97 \\ 
 Class 3 * high iron dose & 0.00 & -0.33 & 0.36 & -0.36 & 0.34 & 0.04 & -0.22 & 0.32 & -0.24 & 0.33 & 0.34 & -0.05 & 0.73 & -0.05 & 0.72 \\ 
  IMD2010:~~~1 &  &  &  &  &  &  &  &  &  &  &  &  &  &  &  \\ 
 ~~~~~~~~~~~~~~~~2 & 0.08 & -0.12 & 0.27 & -0.11 & 0.29 & 0.05 & -0.12 & 0.21 & -0.11 & 0.21 & 0.02 & -0.19 & 0.24 & -0.19 & 0.23 \\ 
  ~~~~~~~~~~~~~~~~3 & -0.05 & -0.25 & 0.16 & -0.27 & 0.15 & 0.08 & -0.08 & 0.23 & -0.07 & 0.23 & -0.09 & -0.31 & 0.12 & -0.31 & 0.12 \\ 
  ~~~~~~~~~~~~~~~~4 & -0.23 & -0.42 & -0.03 & -0.43 & -0.04 & -0.05 & -0.21 & 0.11 & -0.21 & 0.12 & -0.41 & -0.63 & -0.21 & -0.63 & -0.20 \\ 
   ~~~~~~~~~~~~~~~~5 & -0.38 & -0.56 & -0.20 & -0.56 & -0.18 & -0.05 & -0.24 & 0.14 & -0.23 & 0.15 & -0.38 & -0.60 & -0.17 & -0.59 & -0.17 \\ 
  ~~~~~~~~~~~~~~~~NA & -0.01 & -0.20 & 0.16 & -0.19 & 0.18 & -0.11 & -0.28 & 0.06 & -0.28 & 0.06 & -0.22 & -0.43 & -0.02 & -0.42 & -0.02 \\ 
   \hline
\end{tabular}}
\end{sidewaystable}

\section{Simulation Study}\label{simulation}
In this section, we conduct a simulation study to evaluate the performance of  bootstrap methods in constructing confidence intervals for parameters in the discrete-time Markov model for multi-level data.
\subsection{Model for transition probabilities}
 The design of the simulation study is motivated by  the CPRD data analysis reported in  Section 5 of the main text and Section~\ref{analysis}. For simplicity, we only consider two disease states (State 1: no improvement of haemoglobin; State 2:  improvement of haemoglobin) and use logistic models instead of multinomial logistic models for examining the associations between covariates and  transition probabilities.  
 
Within the $j$th course of treatment for the $i$th patient in the $g$th GP practice, the  probabilities of transitions State 1 $\rightarrow$ State 2 and State 2 $\rightarrow$ State 2 follow the logistic models 
\begin{eqnarray}\label{simmodel}
&&\log\left\{\frac{\mbox{Pr}\left(S_{gij,k+1}=2 \mid S_{gijk}=s\right)}{\mbox{Pr}\left(S_{gij,k+1}=1 \mid S_{gijk}=s \right)}\right\} \\
&=&\beta_{0}^{s}+\beta_{1}^{s} \mbox{I}(j=2) +\beta_{2}^{s}  \mbox{I}(j=3) + \beta_{3}^{s} \mbox{I}(j\ge 4)+ \beta_{4}^{s} \log(v_{gijk})\nn \\
&+&\beta_5^s x_{1,gij}+\beta_6^s x_{2,gij}+ \beta_7^s x_{3,gij}+\beta_8^s x_{1,gij}x_{3,gij}+\beta_9^s x_{2,gij}x_{3,gij} \nn\\
&+&b_{g}^{s}+b_{gi}^{s} ~~~~ (s=1, 2). \nn
\end{eqnarray}  Here $v_{gijk}$ is the gap time between the $k$th  and $(k+1)$th visits. $x_{1,gij}$ and 
$x_{2,gij}$ are binary indicators for medium and high daily doses  of oral iron prescribed. $x_{3,gij}$ is the binary indicator for ferrous sulphate (with ferrous fumarate as the reference). $b_{g}^{1}$ and $b_{g}^{2}$ are the GP-practice-level random intercepts, and $b_{gi}^{1}$ and $b_{gi}^{2}$ are the patient-level random intercepts.  We choose the true values of the regression coefficients in~\eqref{simmodel}  based on the parameter estimates from a simplified model for the CPRD data (i.e., some covariates are excluded from the models in Section 3 of the main text). Further, the daily iron dose categories follow a multinomial distribution with $\mbox{Pr}(\mbox{low daily iron dose})=0.1$, $\mbox{Pr}(\mbox{medium daily iron dose})=0.4$ and $\mbox{Pr}(\mbox{high daily iron dose})=0.5$. The binary indicator for ferrous sulphate follows a Bernoulli distribution with $\mbox{Pr}(x_{3,gij}=1)=0.5$. The gap time at the log scale, $\log(v_{gijk})$, follows a standard normal distribution. 

It is assumed that all random effects are independent of  the covariates in the model in~\eqref{simmodel}. Further, the GP-practice-level random intercepts and patient-level random intercepts are mutually independent and we assume that 
$(b_g^{1}, b_g^{2})\trans \stackrel{i.i.d}{\sim} N\left(\mathbf{0}, \left[\begin{array}{cc}
\lambda_{11}^2 & 0  \\
0   & \lambda_{22}^2
\end{array} \right] \right)$,
$(b_{gi}^{1}, b_{gi}^{2})\trans \stackrel{i.i.d}{\sim} N\left(\mathbf{0},\left[\begin{array}{cc}
\sigma_{11}^2 & \rho\sigma_{11}\sigma_{22}  \\
\rho\sigma_{11}\sigma_{22}   & \sigma_{22}^2
\end{array} \right] \right)$, where  $\rho$ is a correlation parameter. In the CPRD data, we would expect that patients who had higher transition probabilities from State 1 (no improvement  of haemoglobin) to State 2 (improvement of haemoglobin) would also be more likely to stay in  State 2. In this scenario,  the patient-level random intercepts in the models for the two transition probabilities would be positively correlated (i.e., $\rho>0$). Therefore, we specify the following  values for the correlation parameter $\rho\in \{0, 0.2,0.6\}$. In addition, the variance parameters are set as $\lambda_{11}=0.65$, $\lambda_{22}=0.80$, $\sigma_{11}=0.85$ and $\sigma_{22}=0.80$. Note that the choice of variance parameter values cannot be informed by the CPRD analysis since it was based on composite likelihood without random effects, but the chosen values reflect reasonable variations at the logit scale. 

\subsection{Data generating steps}
The specific steps to generate the simulated data are as follows:
\begin{enumerate}
\item[1)] For each of the 663 GP practices,  sample the number of patients using the empirical distribution for  the number of patients by GP practices in the CPRD data.
\item[2)] Sample the number of courses of treatment for each of the patients in each of the 663 GP practices using a Geometric distribution with the mean equal to the estimated mean in the empirical distribution for the number of courses of treatment within patients and GP practices in the CPRD data. 
\item[3)] Sample the number of visits within each course of treatment for each of the patients in the GP practices using the corresponding empirical distribution in the CPRD data. Since only $2\%$ of the course of treatment in the CPRD data had more than 6 assessment visits, in the simulated data the number of visits per course of treatment is capped  at 6. 
\item[4)] Sample the random effects $b_{g}^{1}$, $b_{g}^{2}$ for each GP practice, and $b_{gi}^{1}$ and $b_{gi}^{2}$ for each patient within a GP practice. 
\item[5)] Sample $x_{1,gij}$, $x_{2,gij}$ and $x_{3,gij}$ for the $j$th course of treatment of the $i$th patient in the $g$th GP practice.
\item[6)] Assuming that at the beginning of  a course of treatment (i.e. when $k=1$) a patient is always in State 1 (no improvement of haemoglobin), sample the disease states at each  follow-up  visit  sequentially  for the $j$th course of treatment of the $i$th patient in the $g$th GP practice as follows:
\begin{enumerate}
\item Sample the gap time $\log(v_{gijk})$.
\item Conditional on the state at the $k$th visit, the covariates and random effects, use the  logistic model in~\eqref{simmodel} to generate the disease state at the $(k+1)$th visit.
\item Repeat Steps (a)-(b) until $k+1$ equals the number of visits for the $j$th course of treatment for the $i$th patient in the $g$th GP practice.
\end{enumerate}
\item[7)] Repeat Steps 5)-6) for other courses of treatment until $j$ equals the number of courses of treatment for the $i$th patient in the $g$th GP practice. 

\item[8)] Repeat Steps 5)-7) for all patients  in the 663 GP practices.
\end{enumerate}
In total, 800 simulated datasets are generated for each value of the  correlation parameter $\rho\in \{0, 0.2,0.6\}$. 

\subsection{True values for the population-averaged  covariate effects}

Because the data are generated from the logistic model with  multi-level random effects  for the transition probabilities and our composite likelihood approach for estimation provides the estimates of  population-averaged  regression coefficients (i.e., marginal covariate effects), it is necessary to determine the true values of marginal  covariate effects corresponding to the random-effect logistic model in~\eqref{simmodel}. Unfortunately, due to the conditioning on the previous disease state, it is not possible to derive the marginal covariate effects in a closed form in our settings. This is analogous to the situation  where there is no simple relation between  population-averaged  hazard ratios in a marginal Cox model and subject-specific hazard ratios in a Cox model with random effects (i.e., frailties).

To approximate the true marginal covariate effects, we generate a dataset with extreme sample sizes such that the number of GP practices is 100 times of that in the CPRD data and the number of patients within a GP practice is equal to $1000\cdot n_g^{1/4}$, where $n_g$ is the observed number of patients for the $g$th GP practice in the CPRD data.  When $\rho=0$, this dataset consists of observations from 
 549,415,802  visits from 227,644,731 patients from 66,300 GP practices. We then fit a logistic model with the regression structure as in \eqref{simmodel} for the marginal transition probabilities without random effects. The estimated regression coefficients from this dataset are treated as true values of the marginal covariate effects on the transition probabilities.

Finally, we fit the logistic model in~\eqref{simmodel} and obtain point estimates of the regression coefficients. The direct bootstrap and EFB are applied to construct 95\% confidence intervals. We then assess the performance of these bootstrap methods using coverage probabilities of the  95\% confidence intervals.

\subsection{Results}

Table~\ref{simtable} summarizes the bias of the point estimates and coverage of 95$\%$ confidence intervals based on the direct bootstrap and EFB. It appears that for all scenarios with different correlation parameter values $\rho$, the bias of the regression coefficient estimator in the logistic model with composite likelihood is negligible. It is also reassuring to see that confidence intervals from  both direct bootstrap and EFB have good coverages. The correlation parameter seems to have minimal impact on the performance of the point estimator and  the bootstrap methods.

Overall, these results suggest that, in the model selection stage, it might be more  efficient computationally to use the EFB for constructing confidence intervals when analysing  multi-level data of large volume. After the model is finalized, then both bootstrap methods can be applied to provide inference for the parameters of interest. 

\begin{table}[htp]
\caption{Simulation results for comparing the performance of direct bootstrap (DB) and estimating function bootstrap (EFB).}
\label{simtable}
\centering
\begin{tabular}{llrrccrrcc}
  \hline
&  & \multicolumn{4}{c}{\textbf{State 1 $\rightarrow$ State 2 } }&\multicolumn{4}{c}{\textbf{State 2 $\rightarrow$ State 2 } }\\
 
& & True & Bias & \multicolumn{2}{l}{Coverage (\footnotesize{$\%$})}& True & Bias & \multicolumn{2}{l}{Coverage (\footnotesize{$\%$})} \\ 
& & value & (\footnotesize{$10^{-4}$}) & DB & EFB & value & (\footnotesize{$10^{-4}$}) & DB & EFB \\ 
  \hline
$\rho=0$ &Intercept & -0.3175 & -28 & 94.1 & 94.5 & 0.0929 & 3 & 95.0 & 95.2 \\ 
&  $\mbox{I}(j=2)$ & 0.0500 & -2 & 93.8 & 94.1 & 0.0498 & -4 & 94.8 & 94.0 \\ 
&  $\mbox{I}(j=3)$ & 0.0883 & 2 & 94.9 & 94.8 & 0.1293 & 2 & 93.5 & 93.0 \\ 
&  $\mbox{I}(j\ge 4)$ & 0.1053 & -3 & 94.4 & 94.1 & 0.1827 & -7 & 93.2 & 93.4 \\ 
&  $\log(v_{gijk})$ & 0.6172 & 1 & 92.6 & 92.9 & 0.3083 & -6 & 95.0 & 95.4 \\ 
&  $x_{1,gij}$ & 0.1477 & 5 & 94.2 & 94.5 & 0.1572 & 5 & 95.0 & 95.2 \\ 
&  $x_{2,gij}$ & 0.2367 & 5 & 94.9 & 94.6 & 0.0984 & -6 & 95.4 & 95.1 \\ 
&  $x_{3,gij}$ & 0.2055 & 6 & 94.6 & 94.9 & 0.2306 & -30 & 93.2 & 93.2 \\ 
&  $x_{1,gij}x_{3,gij}$ & -0.1296 & -8 & 94.8 & 94.8 & -0.2672 & 22 & 93.5 & 93.0 \\ 
&  $x_{2,gij}x_{3,gij}$ & -0.1594 & 7 & 94.4 & 94.4 & -0.2482 & 38 & 93.6 & 93.9 \\ 
  &&&&&&&&&\\
$\rho=0.2$ &  Intercept & -0.3219 & -32 & 95.0 & 94.4 & 0.1218 & -37 & 96.2 & 95.9 \\ 
&  $\mbox{I}(j=2)$  & 0.0504 & 2 & 94.5 & 93.8 & 0.0491 & -0 & 94.9 & 94.6 \\ 
&  $\mbox{I}(j=3)$ & 0.0887 & 1 & 93.9 & 94.1 & 0.1279 & -1 & 95.0 & 94.5 \\ 
&  $\mbox{I}(j\ge 4)$ & 0.1054 & -2 & 94.5 & 94.8 & 0.1810 & -6 & 95.4 & 95.6 \\ 
&  $\log(v_{gijk})$ & 0.6168 & -2 & 95.0 & 95.5 & 0.3094 & 5 & 94.5 & 93.1 \\ 
&  $x_{1,gij}$ & 0.1476 & 6 & 93.6 & 93.4 & 0.1588 & 15 & 96.0 & 96.0 \\ 
&  $x_{2,gij}$ & 0.2370 & 14 & 92.9 & 93.1 & 0.0995 & 10 & 95.8 & 94.9 \\ 
&  $x_{3,gij}$ & 0.2059 & 23 & 94.8 & 94.4 & 0.2322 & -7 & 95.5 & 96.1 \\ 
&  $x_{1,gij}x_{3,gij}$ & -0.1299 & -18 & 94.8 & 94.8 & -0.2700 & 13 & 96.0 & 95.8 \\ 
&  $x_{2,gij}x_{3,gij}$ & -0.1604 & -28 & 94.5 & 94.1 & -0.2497 & 8 & 95.2 & 95.9 \\ 
    &&&&&&&&&\\
 $\rho=0.6$ &   Intercept & -0.3238 & 2 & 95.4 & 95.6 & 0.1935 & 14 & 95.9 & 95.0 \\ 
  &$\mbox{I}(j=2)$  & 0.0503 & -4 & 93.6 & 93.4 & 0.0467 & 0 & 94.2 & 94.4 \\ 
  &$\mbox{I}(j=3)$ & 0.0882 & -1 & 93.4 & 93.5 & 0.1248 & -7 & 94.1 & 94.6 \\ 
  &$\mbox{I}(j\ge 4)$ & 0.1049 & -4 & 95.6 & 95.0 & 0.1783 & 15 & 94.5 & 93.6 \\ 
  &$\log(v_{gijk})$ & 0.6175 & -1 & 94.1 & 94.4 & 0.3105 & 0 & 94.2 & 94.6 \\ 
  &$x_{1,gij}$ & 0.1480 & 3 & 94.0 & 93.8 & 0.1583 & -2 & 95.1 & 95.4 \\ 
  &$x_{2,gij}$ & 0.2370 & 6 & 94.2 & 94.8 & 0.0996 & 4 & 94.9 & 94.5 \\ 
  &$x_{3,gij}$ & 0.2063 & 9 & 94.9 & 94.1 & 0.2324 & -14 & 95.8 & 95.2 \\ 
  &$x_{1,gij}x_{3,gij}$ & -0.1303 & -4 & 95.2 & 95.0 & -0.2693 & 18 & 96.0 & 95.5 \\ 
  &$x_{2,gij}x_{3,gij}$ & -0.1604 & -10 & 94.5 & 94.9 & -0.2501 & 12 & 95.5 & 95.9 \\ 
   \hline
\end{tabular}

\end{table}